# Emil Rupp, Albert Einstein and the Canal Ray Experiments on Wave-Particle Duality: Scientific Fraud and Theoretical Bias[a]

Jeroen van Dongen[*]

*Institute for History and Foundations of Science*
*Utrecht University*
*PO Box 80.000, 3508 TA Utrecht, the Netherlands*
*&*
*Einstein Papers Project, California Institute of Technology, Pasadena, CA 91125 USA*

**Abstract**

In 1926 Emil Rupp published a number of papers on the interference properties of light emitted by canal ray sources. These articles, particularly one paper that came into being in close collaboration with Albert Einstein, drew quite some attention as they probed the wave versus particle nature of light. They also significantly propelled Rupp's career, even though that from the outset they were highly controversial. This article will review this episode, and in particular Rupp's collaboration with Einstein. Evidence that Rupp forged his results is presented and their critical reception in the socially and politically divided German physics community is discussed. These divisions fail to explain the full dynamic; the latter is attempted by turning to the role that theoretical bias on occasion has in assessing experiment. Einstein's responses in particular are analysed in this context.

## Introduction: the career of Emil Rupp

"[Emil] Rupp, in the late twenties, early thirties, was regarded as *the* most important and most competent experimental physicist. He did incredible things. [...] Later, it turned out that *everything* that he had ever published, everything, was forged. This had gone on for ten years, ten years!"[1] As this quote of Walther Gerlach illustrates, the first third of the twentieth century witnessed one of the biggest scandals in physics: the rise and fall of Emil Rupp.

In 1935 Rupp very publicly retracted no less than five of his scientific publications from the previous year. The articles dealt with such subjects as the polarization of electrons and the artificial production of positrons.[2] Rupp published his retraction in a short notice that appeared in the *Zeitschrift für Physik*. He stated that his withdrawals were the result of an illness and supplied a medical opinion—by a "Dr. E. Freiherr von

---

[a] To appear in: *Historical Studies in the Physical and Biological Sciences* **37** Suppl. (2007), 73-120.

Gebsattel"—in support of his claim:

> Dr. Rupp had been ill since 1932 with an emotional weakness (psychasthenia) linked to psychogenic semiconsciousness. During this illness, and under its influence, he has, without being himself conscious of it, published papers on physical phenomena [...] that have the character of 'fictions.' It is a matter of the intrusion of dreamlike states into the area of his scientific activity.[3]

After years of producing highly controversial work in Heidelberg, Göttingen and now Berlin, Rupp had finally overplayed his hand: he had claimed that in his production of positrons, a beam of protons had been accelerated at potential differences of 500 kV. For this he obviously lacked the necessary apparatus or the means to acquire such an apparatus—he even did not have enough laboratory space to accommodate such an accelerator. Some of his colleagues at the Berlin research laboratory of the *Allgemeine Elektrizitäts Gesellschaft* (AEG), in particular Arno Brasch and Fritz Lange, had grown suspicious after hearing of Rupp's latest claims. They confronted Rupp and his supervisors and this led the AEG to draw up an internal report that was most damaging for Rupp. Subsequently Carl Ramsauer, the director of the AEG laboratory, had dismissed him.[4]

Ramsauer had further felt it necessary, in an attempt to control the damage done to the reputation of his institute, to publish a statement of his own in the *Zeitschrift für Physik*. Rupp had made clear that in his opinion, "there exists no reason to retract earlier works either wholly or partially." Ramsauer, however, went out of his way to state in print that Rupp's earlier work, too, was not to be trusted. He particularly warned for "Rupp's papers on canal rays, which have been questioned over and over again."[5]

Traces of Rupp's earlier work can easily be found in the contemporary literature. One only needs to look in Werner Heisenberg's 1929 Chicago lectures on "The Physical Principles of the Quantum Theory" to become convinced of Rupp's prestige. When Heisenberg discussed seven "Important Experiments," he included, along with the Davisson-Germer experiment on the diffraction of matter and the Compton experiment, an experiment that he termed "The Experiment of Einstein and Rupp."[6] Indeed, Rupp's controversial work on canal rays was carried out in 1926 in close collaboration with Albert Einstein. Did, as is strongly suggested by the words of Ramsauer, Rupp truly forge his results in these experiments? If so, how and why? What was Einstein's role?

And how did the physics community respond to Rupp's claims?

This paper will present strong documentary evidence suggesting that the canal ray results were indeed forged, as an important article by Anthony P. French[7] has also argued. However, where French has given a general overview of Rupp's career, the current paper will delve deeper into the details of Rupp's canal ray work and address broader issues of scientific practice that their reception raises; in particular, the relation between theory and experiment, foremost in the context of Einstein's views on physics, will be discussed.

The article will start by outlining Rupp's first experimental work on canal rays. It will focus on Einstein's interest in this work and his ensuing collaboration with Rupp. It follows the unfolding controversy—for to some contemporaries, Rupp's work immediately appeared to be incomprehensible and even suspect, and these suspicions were very visibly published. Yet, the repercussions for Rupp seem to have been coming rather slowly; he published controversial work on leading subjects for at least nine years, while his career in physics advanced. His association with Einstein may have worked to his benefit; many might have accepted his fraudulent results because of Einstein's endorsement of his experiments.[8] Moreover, some aspects of Rupp's work were quite convincing, and—as will be shown here—he was rather agile in dealing with the criticisms. Ultimately, one of science's safeguard mechanisms came into play: replication.[9] A circle of Munich experimentalists around Wilhelm Wien and then Walther Gerlach directly challenged Rupp on this score.

One of the most intriguing questions of this episode remains why such renowned and very able physicists, who certainly could be very critical of any of their colleagues' work—men such as Max von Laue and, in particular, Einstein—engaged with Rupp and his experiments as much as they did. To understand this better this article will, in the concluding sections, turn to social and political divisions present in the German physics community. These can shed some light on the case of Rupp's canal ray experiments, but fail to explain the full dynamic; the latter is best achieved when involving the familiar role that theoretical bias can have in assessing experiment. A companion publication outlines the role that Rupp's experiments played in the history of quantum theory.[b]

---

[b] Jeroen van Dongen, "The interpretation of the Einstein-Rupp experiments and their influence on the history of quantum mechanics," *Historical Studies in the Physical and Biological Sciences* **37** Suppl. (2007) 121-131.

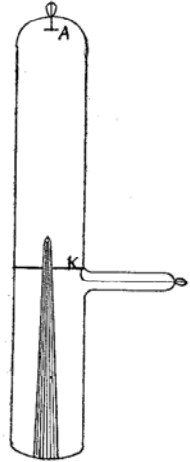

Figure 1: Canal rays, moving through the hole in cathode *K* and into the lower chamber, in which there is no electric field and pressure is very low. The canal ray light is radiated out in all directions. Taken from Wien 1927 (note 13), p. 436.

## Canal ray light and its interference properties

Emil Rupp was born in 1898 in Reihen, in the German state of Baden. He studied at the University of Heidelberg and took his doctorate under Philipp Lenard, submitting a thesis that dealt with the absorption properties of phosphors and receiving the Ph.D. *summa cum laude*.[10] Rupp stayed on in Heidelberg to work on his *Habilitation* and succeeded in acquiring the permission to teach in early 1926. His *Habilitation* thesis, which was published in the *Annalen der Physik* and by Rupp often referred to as his "*Annalenarbeit*," was a detailed study of light emitted by canal ray sources, in particular of the interference properties of such light.[11]

Canal rays are formed in gas discharges between an anode and cathode, and are seen most clearly when a hole is made in the cathode; one sees a lightened up beam propagate through the hole (or "canal") into the vacuum chamber behind the cathode, where no electrical field is present (see figure 1). These canal rays were discovered in 1886 by Eugen Goldstein, but following their discovery they remained largely unstudied for some eleven years, until Wilhelm Wien turned his attention to the subject.[12] Wien's sudden interest may have been sparked by the recent discovery of X-rays; one believed that canal rays, too, were possibly some fundamental form of radiation. Wien's many and elaborate experiments, done over a period of some thirty years in both Würzburg and Munich, eventually revealed that canal rays actually consist of ions and atoms, constantly losing and regaining charges and emitting light while in the gas discharge tube. After exiting through the cathode's canal, into the vacuum chamber, a beam of canal rays would radiate over a short distance until it had released all its energy. Once the beam had left the high pressures of the discharge chamber, the canal ray source was considered to be very well suited for studying the undisturbed, spontaneous emission of light.[13]

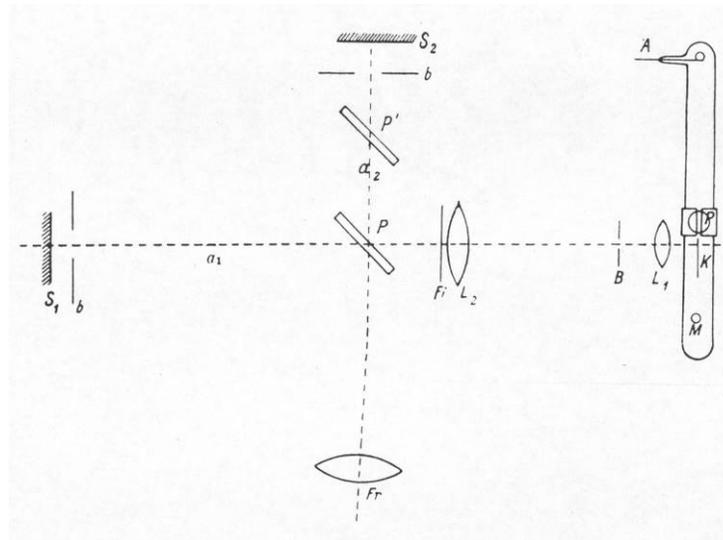

Figure 2: Rupp's original experiment, taken from Rupp 1926a (note 11). Lens $L_1$ will be omitted in the remainder of our article. Light, emitted at $K$, travels to the left in this picture.

In the 1920's, Wien—by then the world's leading authority on the subject—introduced two time scales in his studies of canal rays as light sources. The retention time (*Verweilzeit*) was a measure for the time that the atom remained in its excited state before emitting a light quantum, while the damping time (*Abklingzeit*) was a measure for the decrease of amplitude of an emitted lightwave.[14] Obviously, the concept of retention time was more closely linked to the atomic models of Bohr and Sommerfeld—in which light is emitted as an electron goes from one Bohr orbit to another—whereas the damping time could more easily be identified with the waves and oscillators perspective of classical theory.

Wien held that models that employed either concept could describe equally well the observed decreasing intensity of the emitted light along the beam behind the cathode: a beam that was only visible over a short distance consisted of atoms that could have a short retention time, or a short damping time. In experiments done in 1925, he further found that only one of these time scales ought to be relevant: either the damping time was negligible relative to the retention time, or vice versa. It was this issue, an issue that probed quantum and classical emission models, which had inspired Rupp's initial experiments on the interference properties of canal ray light: by determining the maximum value for the coherence length of the emitted light, one could calculate a lower bound for the damping time. [15]

The most important result of Rupp's *Annalenarbeit* was thus the measurement of the maximum coherence length of light emitted by hydrogen (at the $H_\beta$ spectral line, $\lambda = 4861\,\text{Å}$) and mercury (at $\lambda = 5461\,\text{Å}$) canal rays. To obtain these results, he used the set-up with a Michelson interferometer reproduced in figure 2.[16] The canal ray beam $K$ passes in front of lens $L_1$ that produces an unmagnified image of the beam at slit $B$ (lens $L_1$ will be dropped from our discussion of this experiment in the remainder). This slit is in the focal plane of a second lens $L_2$. The canal ray light, after passing through a filter $Fi$ that should select the correct spectral line, then propagates to the interferometer. After beam splitter $P$, it falls on either mirror $S_1$ at a distance $a_1$ from $P$ or mirror $S_2$ at a distance $a_2$ ($P'$ is a compensator). In front of both mirrors are additional slits of unspecified width, named $b$. The light is recombined at lens $Fr$ that projects an interference pattern on a screen that is not in the diagram. Finally, since $a_1$ is greater than $a_2$, Rupp defined the path difference picked up in the interferometer as $\Delta l = 2(a_1 - a_2)$.

Rupp observed the interferences with his bare eyes ("subjective observation"[17]), even though the source was quite weak. He pointed out that this could lead to errors, and therefore limited his observation time to at most half an hour. He observed the interference pattern and moved one of the interferometer's mirrors until the pattern could no longer be seen for a maximum distance between the mirrors; this would constitute the observation of the coherence length. Rupp further stated that the inclusion of the slits $b$ brought out the interference fringes most clearly.

This last claim becomes clearer when considering the reasons that Rupp gave for the destruction of the interference that would occur before the maximum coherence length had actually been reached. Among those reasons were the possible multiplet structure of the spectral lines, collisions in the beam and disruptive electrical fields. But these were not all—the most interesting reasons that Rupp listed were the *Doppler shifts in the emitted light* due to the *velocity of the beam*, and further Doppler shifts due to the *thermal motion of the atoms* in the beam.[18] The beam motion Doppler effect would smear out the wavelength and thus reduce the observed coherence length (which is proportional to $1/\Delta\lambda$).[19] The thermal motion would lead to different wavelengths and different fringe patterns for the individual atoms, and thus prevent the constructive formation of an overall pattern—again the maximum path difference at which interference could still be seen is reduced. These two disturbances would later lead to great concern.

Rupp himself did not say too much about the problems that the thermal motion could produce; he gave a formula for the spectral line width due to the thermal motion and remarked that because of its lighter mass, the line width for hydrogen is broader than for mercury—and thus its maximum coherence length will be shorter.[20] He however did not further quantify or elaborate on how restrictive the thermal motion would be.

He was a bit more expansive when it came to describing the consequences of the Doppler effect due to the beam motion; Rupp pointed out that the beam's velocity will in general have a component in the direction of the emitted radiation, and thus indeed lead to a broadening of the wavelength. As a measure for the magnitude of this spreading, he introduced an angle $\varphi$ with respect to the normal for the incident radiation that by slit *B* was limited to 20'. The definition of $\varphi$ was rather ambiguous however (it is for instance not clearly stated which of the two lenses $L_1$ or $L_2$ Rupp used in his definition of $\varphi$) yet this was all that was said about the disruptive influence of the beam motion Doppler effect. Rupp believed that further limiting the radiation field by more slits (*b* in his figure) should limit the Doppler components enough to prevent them from impeding a good result on the maximum coherence length, but he did not give a value for the reduction by the slits *b*. Finally, he stated that the Doppler shifts could be increased due to diffusion as the beam leaves the canal. Yet, in Rupp's opinion this diffusion effect should be negligible since not many atoms ought to undergo a sideways motion.[21]

The results that Rupp obtained were amazing. For the *Hg* 5461 Å-spectral line he found a maximum coherence length of 62 cm, just three centimeters less than the value that had earlier been found by Ernst Gehrcke and Otto Lummer (whose work he cited) for the same source at rest, *without* the beam motion.[22] The result for the $H_\beta$ spectral line was even more impressive: here Rupp had found a maximum coherence length of no less than 15.2 cm—well above the maximum length for an ordinary hydrogen source which was expected to be 3.5 cm.[23] Rupp's lengths had reportedly been obtained for various widths of slits and beam and at various speeds and potential differences.[24]

These strong results led Rupp to further conclude that, under the assumption that only light originating from the same atom can interfere, there was a lower bound for the damping time given by the maximum coherence length divided by the speed of light: $0.5 \cdot 10^{-9}$ s for $H_\beta$ and $2 \cdot 10^{-9}$ s for *Hg*. Surprisingly, he however would not draw any conclusions on the existence of a retention time: Rupp believed that also in the quantum picture of emission, light should still somehow consist of a finite wave train. This wave would then still account for the observed coherence length, and he would thus not rule out the concept of retention time.[25]

Rupp's *Annalenarbeit* created quite a stir. By some it was received favorably; in fact, shortly after the paper came out he was offered a position as Robert Pohl's assistant at the First Physical Institute of the prestigious University of Göttingen, which he accepted.[26] Reviews of Rupp's work also soon appeared. Eduard Rüchardt of Munich, a former student of Wien and now working his way up as the next authority on canal rays, wrote an abstract for the *Physikalische Berichte* that was quite critical. He pointed out that Rupp's vacuum pump appeared to be in the wrong location; obtaining the kind of freely decaying atoms that Rupp had wanted to do his experiments with should in fact have been impossible.[27] But more importantly, an elaborate article by the astronomer Walter Grotrian, in a highly visible journal, *Die Naturwissenschaften*, discussed the results positively.[28] Grotrian further drew Albert Einstein's attention to Rupp's work.

## Einstein on instantaneous light emission

After having learned of Rupp's work, Einstein himself soon submitted a short note to *Die Naturwissenschaften* on 16 March 1926 in which he proposed to do an experiment that should decide whether the emission of light occurred by a process that was extended in time, as in classical theory, or whether it was instantaneous. Of course, an instantaneous emission would provide strong evidence for his own light quantum hypothesis of

1905.[29]

Earlier, in 1921, Einstein had already proposed an experiment that should probe this question— also this experiment used canal rays as a light source.[30] In this case, a light wave was expected to rotate in a dispersive medium due to Doppler shifts present because of the motion of the canal ray beam; instantaneously emitted quanta were somehow supposed not to exhibit such a rotation. Hans Geiger and Walther Bothe performed the experiment and initially appeared to confirm Einstein's intuitions—these intuitions however turned out to have been false: Einstein had made a theoretical mistake when applying the classical wave theory, as he quickly admitted himself.[31] He had overlooked that in the analysis he should have studied a wave packet instead of an infinitely long wave train. As Paul Ehrenfest had explained to him, no rotation or deflection was to be expected if the wave packet was properly accounted for and the classical and quantum predictions indeed turned out to be the same.

Nevertheless, by 1926 the light quantum hypothesis had received strong experimental support, most notably through the 1923 discovery of the Compton effect.[32] Furthermore, in 1925 Bothe and Geiger showed that the Bohr-Kramers-Slater theory, much maligned by Einstein and a last attempt to maintain a full radial wave picture of light emission, could not hold.[33] Finally, in the early months of 1926 Walther Bothe once more submitted results that strongly supported Einstein's light quantum: Bothe showed that absorption events occurred in an entirely uncorrelated fashion, which posed another problem for the idea that light was radiated out in radial waves.[34]

Einstein in March of 1926 must thus have been pretty confident that the concept of a particulate, instantaneous emission would also find experimental confirmation. The wave character of light, Einstein by then believed, should not be due to a temporally extended process of the emitting atom or electron. He did not know a straightforward answer to tough questions on how the wave train—that clearly exhibited itself in interference phenomena—could be brought in line with an instantaneous emission process, yet he expected the latter to be affirmed by his new experiment. The interpretations that Einstein may have entertained of the experiment are addressed separately in the companion piece to the present article—here, his proposed experiment[35] and the ensuing collaboration with Rupp are discussed.

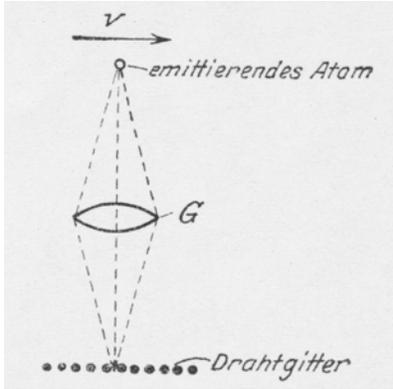

Figure 3: Canal ray atom emitting light and passing along a wire grid. Taken from Einstein 1926a (note 29). The lens *G* will be omitted in the remainder of our article.

The experiment that Einstein sketched was surprisingly simple (it will be called his 'Wire Grid Experiment' [*Gitterversuch*]). Einstein imagined an emitting atom that passed with velocity *v* along a wire grid or grating. The light emitted was to be focussed by a lens *G* on the grid, such that a sharp image of the atom is formed in the plane of the grid (see figure 3). Einstein took a grid in which the openings are $b = 0.1$ mm wide,[36] and the distance separating the center of two openings is *2b*.

If light is not emitted instantaneously, the wave train will be cut up in pieces, each of length $c\tau = cb/v$ and a distance $c\tau = cb/v$ apart (see figure 4). The atom is only visible, that is, when it passes behind a slit in the wire grid, and it takes a time $\tau = b/v$ to pass a slit. However, if light is emitted instantaneously, then the motion along the grid should have no effect: should a radiating atom be passing behind a slit, somehow the entire wave train would get transmitted.

The interference properties of the light coming out of the grid were to be studied by using a Michelson interferometer. If the classical picture would apply, the visibility of the interference pattern should vary with the path difference introduced by the interferometer. Interference should then completely disappear for values of the path difference equal to

$$(2n+1) \times cb/v, \qquad (1)$$

with *n* an integer. For those values, the transmitted part of the original wave train would have to interfere with the non-transmitted part; so there would not be any interference. The interference should be best visible when the path difference is

$$2n \times cb/v. \qquad (2)$$

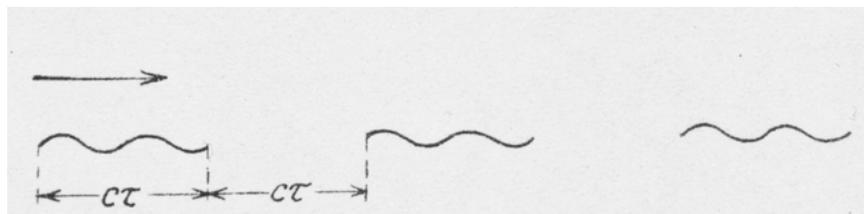

Figure 4: In the classical picture, the emitted light is cut up as the atom moves along a grid. Taken from Einstein 1926a (note 29).

If the classical picture would not apply however—that is, if light would be emitted instantaneously— then this variation should not be observed, since somehow the wave train would get transmitted in full. This would imply that "the interference properties of the radiation have no relation to any periodicity of the emitting atom."[37] Einstein did not explicitly say so in his article in *Die Naturwissenschaften*, but it is clear that this is what he initially expected to be the outcome; he wrote to Erwin Schrödinger in April 1926 that he had "found arguments that pretty much rule out elementary radial waves. I am pretty sure that the experiment outlined by me will be negative."[38]

This 'Wire Grid Experiment' could obviously only be done properly if one could study interference patterns for path differences in excess of $cb/v$. This was where Rupp's *Annalenarbeit* proved its fundamental value: a quick calculation learned Einstein that the wave train intervals for Rupp's fast moving hydrogen atoms would be about 6 cm. Einstein pointed out that Rupp had observed interference at more than twice that value, so he expected that the experiment should certainly be feasible.

Einstein's short note received considerable attention. Paul Ehrenfest soon wrote to him, and rather jokingly remarked that this time (unlike in 1921) Einstein had not made any mistake when applying the wave theory. Yet contrary to Einstein, Ehrenfest expected the classical theory to be confirmed.[39] Another early respondent was the experimental physicist Georg Joos from Jena, who presented his own argument for why the classical theory was likely to be confirmed in Einstein's experiment. Joos wanted his argument to be published in the widely read *Naturwissenschaften*, but despite Einstein's appreciation of his idea, he had to settle for a publication in the *Physikalische Zeitschrift*.[40]

Another commentator, the British spectroscopist Robert d'Escourt Atkinson, did get the opportunity to share his views with the audience of *Die Naturwissenschaften*. The tone of his note was much more critical than Joos'—not of Einstein however, but of Rupp. Atkinson strongly doubted whether Einstein's experiment could be carried out because he could not believe that Rupp had observed interference for the $H_\beta$ spectral line at 15.2 cm path difference. He reminded the reader that for a hydrogen source at rest and at room temperature, *thermal motion leads to Doppler shifts in the emitted light*; this thermal motion surely had to be present in the canal ray beam too. Rupp had pointed this out as well, but Atkinson argued, it meant that the coherence length of the canal ray light could be at most 3.5 cm; the maximum value to be expected for hydrogen at rest.[41] The fact that the atoms in the beam had been accelerated in one direction could have no positive effect on the maximum value of the coherence length, as it could not diminish the thermal motion of the atoms. Atkinson believed that the canal ray value should even be smaller, because one had to expect an increase in the atomic collisions to occur due to the production of the beam. These limitations had made him decide to give up any attempt to determine the light emission time from the path differences of canal ray interferences a few years earlier. He found it very hard to believe that Rupp had managed to steer clear of such collisions, and that in his case even the thermal motion had somehow been cancelled.

Yet, even if that were so, then still Atkinson had an objection. Even if Rupp had such incredibly homogeneous canal rays at his disposal—homogeneous to the degree that they could be regarded as effectively monochromatic sources—then *the Doppler effect due to the beam motion* would further have had to strongly limit the coherence length. Namely, light waves emitted at an angle to the principal axis of the interferometer would have undergone a Doppler shift due to the motion of the emitting atom. The shift in wavelength ($\Delta\lambda$) due to this Doppler effect was given by:

$$\Delta\lambda = -\lambda_0 \frac{v}{c}\sin\alpha, \tag{3}$$

with $\lambda_0$ the wavelength with the sources at rest, $v$ the source velocity, $c$ the speed of light and $\alpha$ the angle of the emitted light with the *x*-axis (see figure 5).

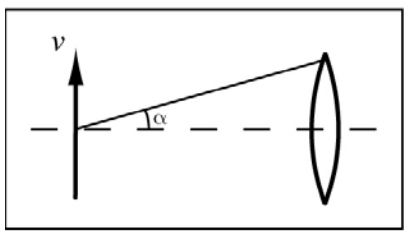

Figure 5: Doppler effect due to beam motion *v*.

As pointed out before, Rupp had also briefly mentioned in the *Annalenarbeit* that this Doppler effect limited the maximum coherence length. Yet, Atkinson derived that if a coherence length of 15.2 cm had actually been reached, then $\alpha$ (and not Rupp's confusing $\varphi$) could not exceed $\alpha = 1.2'$. He noted that indeed Rupp had additional slits *b* located near the mirrors of the interferometer (see figure 2), but insisted that if these had limited $\alpha$ to $1.2'$, then the intensity of the light would in fact have been much too weak to observe the effects claimed by Rupp. In short, Atkinson argued that because of the thermal motion, Rupp could not have been able to see interferences at more than 3.5 cm, and that actually, because of the beam motion, the observed maximum coherence length should be much less at best.

Atkinson's article had been prompted by Einstein's short note and came out in *Die Naturwissenschaften* on 18 June 1926, three months after Einstein's original paper in the same journal. As it appeared in such a visible journal, Einstein may have learned of its contents too. By the time the article was published however, he already had thoroughly thought through the consequences of the Doppler effect of the beam motion and had revised his earlier arguments. Einstein proposed a way in which this effect could be circumvented, and Rupp subsequently claimed to have unknowingly satisfied Einstein's new demand in his earlier work. However, a strong ground for suspicion for critics of Rupp—and an incomprehensible circumstance for those that believed in the veracity of his results—was still how he could have cancelled the other Doppler effect, that due to the thermal motion.

## Einstein and Rupp: first discussions

Einstein had contacted Rupp on March 20, 1926, well before Atkinson's article came out. He suggested to Rupp to do the Wire Grid Experiment and sent him a copy of his note in the *Naturwissenschaften*. Einstein had just begun to rethink the experiments and could not understand one aspect of Rupp's original publication: how indeed had he managed to suppress the *beam motion Doppler effect*? After expressing his confusion over Rupp's angle $\varphi$, he told Rupp that this Doppler effect should lead to "disappearance of the interference at really small path differences. It is a mystery to me how you were able to avoid this effect." He asked whether Rupp had used slits to limit $\alpha$ and in this way had limited the effect's consequences.[42]

Rupp replied promptly: on March 23rd, he informed Einstein that he would gladly do his experiment. Regarding the destructive Doppler shifts Rupp pointed Einstein to the slits in figure 2 but admitted that he did not know the exact value of the effective $\alpha$ in his experiment. He had established his maximum coherence lengths "purely empirically, by moving the slits and lenses around."[43] Einstein was delighted that Rupp was willing to do his experiment; he suggested a joint publication, but was careful to remind Rupp that perhaps the latter's superior, the Heidelberg professor Philipp Lenard would take offense. Lenard was, after all, a fervent anti-relativist—and anti-Semite.[44]

Einstein must not have been satisfied with Rupp's explanation as he continued to think about the consequences of the beam motion Doppler effect. He came to some interesting conclusions that some months later he published in a second theoretical study of the subject (the article came out in the October 21, 1926, issue of the Proceedings of the Berlin Academy).[45] In the case of his own experiment with a grid, there would actually be no adverse consequence of the beam motion. Einstein realized this already when he was writing Rupp on March 31, but in that letter he did not elaborate on the issue. In his later article, that was completely drawn up in manuscript form by May 9, 1926,[46] he explained why.

In the article he first gave the example of a canal ray beam source directly in front of a Michelson interferometer (see figure 6), without either a lens (in Rupp's original arrangement [see figure 2] the relevant lens $L_2$ was included) or a wire grid (as in his own experiment). Einstein argued that because of the symmetries in the problem, one could just as well imagine the canal ray beam at an infinite distance from the interferometer and replace its atoms by imaginary sources at rest, each emitting plane waves with frequencies given by $\lambda = \lambda_0 + \Delta\lambda$, with $\Delta\lambda$ given by relation (3).

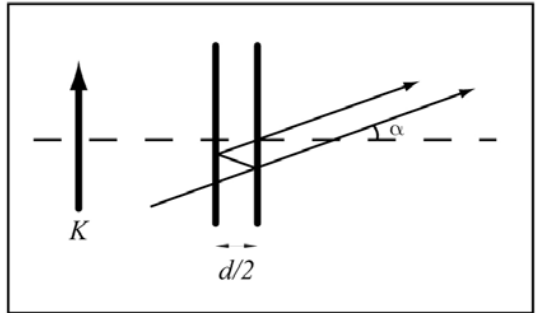

Figure 6: Einstein's first example in Einstein 1926b (note 45): Canal ray source $K$ and the interferometer with distance $d/2$ separating the mirrors (to simplify the diagram, the Michelson interferometer is replaced with a dielectric slab with partially reflecting surfaces; see Klein 1986 [note 47], pp. 288-292). The incident light makes an angle $\alpha$ with the normal.

In a Michelson interferometer, the phase difference between two light waves incident at an angle $\alpha$ from the normal and recombining in a particular fringe is[47]

$$\delta = 2\pi \frac{d}{\lambda} \cos\alpha. \tag{4}$$

For small $\alpha$, the relative phase difference becomes $\delta \approx 2\pi \frac{d}{\lambda}(1 - \frac{1}{2}\alpha^2)$. Taking the Doppler shift (3) into account changes the above to:

$$\delta = 2\pi \frac{d}{\lambda_0(1 - \frac{v}{c}\sin\alpha)} \cos\alpha, \tag{5}$$

which for small $\alpha$ gives $\delta \approx 2\pi \frac{d}{\lambda_0}(1 - \frac{1}{2}(\alpha - \frac{v}{c})^2)$, i.e. the beam motion would only lead to a slight shift of the interference pattern of the order of $v/c$. The conclusion is thus, as each wavelength corresponds to a particular angle and a particular fringe, one should still see an interference pattern, despite the beam motion Doppler effect. This analysis carried over to the situation when a grid would be placed in front of the interferometer, as in Einstein's Wire Grid Experiment. In the mentioned second publication, Einstein re-derived for what values of the path difference one should best see interferences, taking full account of the Doppler effect (3) and diffraction at the grid, and he re-confirmed formulas (1) and (2).

Einstein had also thought about the consequence of the Doppler effect in the case of Rupp's *original* arrangement (figure 2) with a lens in front of the interferometer. ("I would like to know about the interferences in the case of *your* arrangement."[48]) This yielded an interesting and quite surprising conclusion. As said, Einstein must surely have wanted to see how it had been possible for Rupp to have observed interferences in his set-up despite the beam motion Doppler effect, for Rupp's explanation (the slits *b*) was far from convincing. He proposed an ingenious solution: one of the interferometer's mirrors had to be rotated through a small angle in order to get interference. The experiment in the German literature was subsequently known as the *Spiegeldrehversuch*, or 'Rotated Mirror Experiment.' [49]

The essential idea of the 'Rotated Mirror Experiment' was contained in Einstein's letter of March 31st, but again Einstein's later theoretical publication contains the fullest account.[50] A lens with focal length *f* is placed between the canal ray and the interferometer. In the focal plane of the lens will then be formed a particular image of the canal ray—all light originally incident at an angle $\alpha = y/f$ will be collected at height *y* in this image ($B_1$). One can then think of this image as an extended source at rest in which each point at height *y* corresponds to a point source emitting light with wave length $\lambda = \lambda_0 (1 - \frac{v}{c} \sin \alpha) \approx \lambda_0 (1 - \frac{v}{c} \frac{y}{f})$. The mirrors of the interferometer further produce a second, identical image ($B_2$) at a distance -*d* from the first image. Points at the same height *y* in the two images can be taken to be coherent point sources, emitting light at the same wavelength $\lambda$ (see figure 7).

In this arrangement, no clear interference pattern will be formed. As each pair of points in $B_1$ and $B_2$ emits light in every direction to the interferometer, the two coherent points at height *y* produce a complete interference pattern. Yet, because different values of *y* correspond to different wavelengths, each such pair of points forms a different pattern; this obstructs the formation of a clear overall pattern in the focal plane of the interferometer's lens.

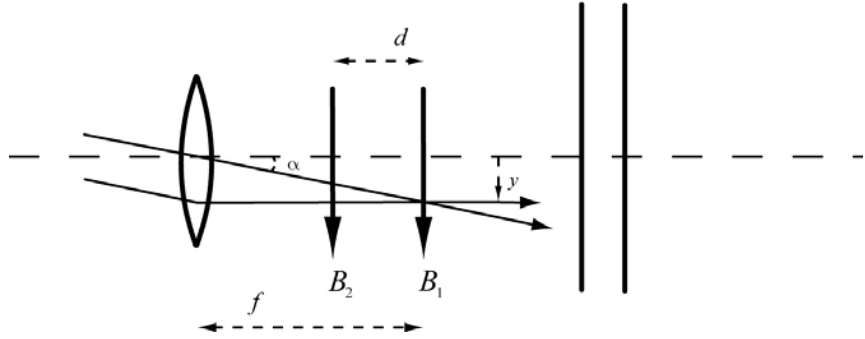

Figure 7: With a lens in front of the interferometer and the canal ray source at infinity, an image $B_1$ of the canal ray source is formed at the focal distance $f$. Image $B_2$ is formed in the reflection in the interferometer's mirrors.

Einstein realized, however, how one can correct for this. He pointed out that if one of the mirrors of the interferometer were rotated through a particular angle, the phase difference between interfering light waves could be tuned in such a way that it no longer depended on $y$. The rotation of one of the mirrors of the interferometer would thus correct for the destructive Doppler effect of the beam motion. Specifically, a rotation of the mirror through an angle $\beta/2$ changes the path difference, measured in wavelength $\lambda$ between the light emitted by the two coherent points at height $y$:

$$\frac{d}{\lambda_0(1-\frac{v\,y}{c\,f})} \rightarrow \frac{d-\beta y}{\lambda_0(1-\frac{v\,y}{c\,f})}, \tag{6}$$

as the picked up path difference between the mirrors grows with double the distance separating them. If one then chooses

$$\beta = \frac{v}{c}\frac{d}{f}, \tag{7}$$

the second expression in (6) is no longer dependent on $y$. This means that the phase difference between light emitted by a pair of coherent sources at all $y$ is again given by (4) with $\lambda_0$, and each pair again produces exactly the same fringes. So a tiny rotation (of the order of $v/c$) of the mirror produces a clear interference pattern.

Einstein argued that also this experiment would decide if light emission would be an instantaneous process or not. To explain this, he again put the canal ray source in the focal plane of the lens (as in Rupp's original *Annalen* experiment) and observed that only those rays can interfere that hit the screen behind the interferometer's lens at the same time and in the same point. Tracing these light rays back to the beam one sees that they have been emitted at different times—if interference nonetheless occurs, one can conclude that they have been emitted by the same atom while it moves in the canal ray beam (see figure 8). Einstein further determined that the two different points in the beam are $\beta f = \frac{v}{c} d$ apart. Would the emission be instantaneous, there would be no interference, since there could be no coherence between light emitted at various locations in the beam. Einstein must thus have come to realize that if in the *Spiegeldrehversuch* interference was observed, this would strongly support the classical emission picture.

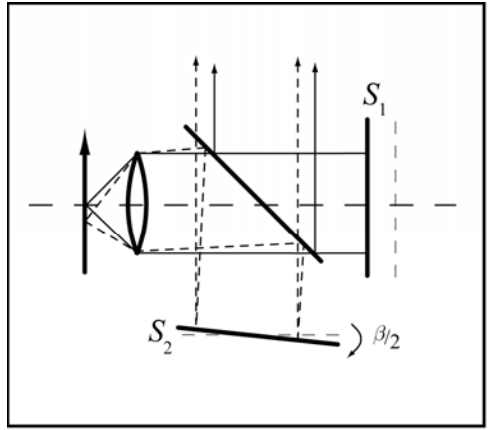

Figure 8: The *Spiegeldrehversuch* ('Rotated Mirror Experiment'): a clear interference pattern appears only after rotating the mirrror, as the rotation changes the path difference in a way that exactly compensates for the blue-and redshifting due to the beam motion Doppler effect. The two interfering signals are emitted at different locations in the beam, meaning that the pattern is made up of light emitted by atoms as they move in the beam.

The arrangement of Rupp's original *Annalen* experiment was equivalent to the *Spiegeldrehversuch* as outlined by Einstein. However, Rupp had not mentioned anything about having rotated a mirror. He nevertheless claimed to have seen interferences: Einstein must have concluded that he had unknowingly rotated a mirror a tiny bit and thereby compensated for the beam motion Doppler effect. On that account, Rupp would already have confirmed the classical picture. Indeed, Einstein wrote to Paul Ehrenfest in early April that "Rupp should do an experiment about [the wave versus particle nature of light]. He has probably already done it,

but—he does not know it yet."[51] Einstein began to expect the classical outcome of his experiments; one of the reasons for his changing position likely was that that outcome had inadvertently already been corroborated by Rupp.[52]

On April 11, 1926, Rupp replied to Einstein's latest letter (that of March 31) in which the latter had outlined the Wire Grid Experiment and the *Spiegeldrehversuch*. He would soon begin to undertake these, Rupp wrote, and would be glad to come to a joint publication; Lenard's attitude towards Einstein would not impede their collaboration as Rupp had all the required instruments already at hand. Einstein, writing to Rupp on April 18, was glad to hear that they were to do this project together—"if it were not the Heidelberg laboratory, I would come over, as the results will be important for the theory of radiation."[53]

## Rupp reports results for the Wire Grid Experiment

Einstein never did visit Rupp in Heidelberg and never saw him do his experiments. Over a period of some two months Rupp sent Einstein his results, leading the latter to comment on these.[54] Rupp sent a first set of preliminary data of the experiment with the wire grid on April 29. In his initial trial with the $Hg$ $\lambda = 5461$ Å line, he had used canal ray atoms moving at a speed of $v = 2 \cdot 10^5$ m/s and a grid with 100 "parts/cm" ["Teile/cm"], i.e. $b = 0.005$ cm. Rupp found that: "interferences stay clear up to [a path difference of] 7 cm, their sharpness declines to a min at 15 cm, increases again to 25 cm, is clear at 30 cm, declines again near 35 cm and is again very unclear at 45 cm."[55] When putting these values into Einstein's formulae (1) and (2), one quickly learns that Rupp's results for the distance separating the maxima of the interference visibility are actually off by a factor of two. Or did he use a different definition for the grid widths $b$ than Einstein? In any case, these were just preliminary results, so Rupp should not be judged by them. He did however re-confirm the results in another note on May 1.[56]

Einstein wrote Rupp again on May 5th. He intended to give some last minute advise—in the case of the *Spiegeldrehversuch*, Einstein once more pointed out to Rupp, who had not yet caught on, that interference was to come about by rotating one of the interferometer's mirrors, and not because of the slits in front of them that Rupp still kept pointing to (for the latter still believed that slits were needed to suppress the beam motion

Doppler effect). Einstein now also strongly stated that "it is as good as certain that everything must come out according to the wave theory. The expectation I expressed in my note [in *Die Naturwissenschaften*] can hardly be correct."[57] Rupp could no longer doubt that he was expected to confirm the classical theory.

Einstein sent the draft of his second detailed theoretical paper to Rupp on May 9, and remarked that this contained "something new, also for you, namely in the case when a lens is placed in front of the interferometer."[58] This comment referred to Rupp's original set-up of the *Annalenarbeit*; perhaps Einstein was here being critical of the fact that Rupp had not mentioned anything about a mirror rotation in his earlier publication—or he just wanted to draw Rupp's attention yet again to the theory of the *Spiegeldrehversuch*, afraid that perhaps the latter had still not understood it. In any case, Einstein further informed Rupp that he now thought it best if they published their work separately, but with both papers immediately following each other.

On May 14 Rupp sent Einstein results of the Wire Grid Experiment, that he had brought to a "certain conclusion";[59] for *Hg* canal ray atoms moving at $1.9 \cdot 10^5$ m/s, he found for a grid with "0.1 mm distance" the following visibility of the interference, as a function of the path difference:

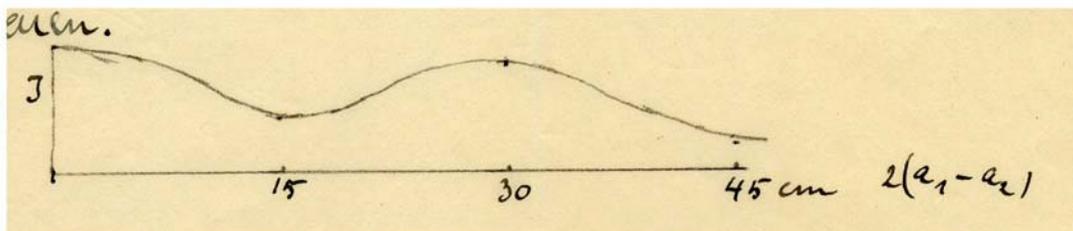

For a wire grid that was twice as wide, Rupp found:

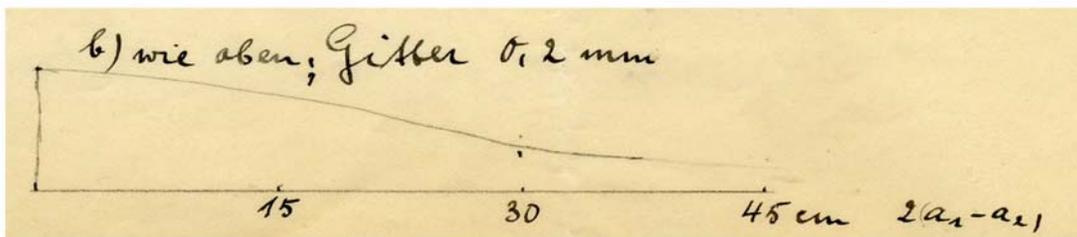

and for a finer grid:

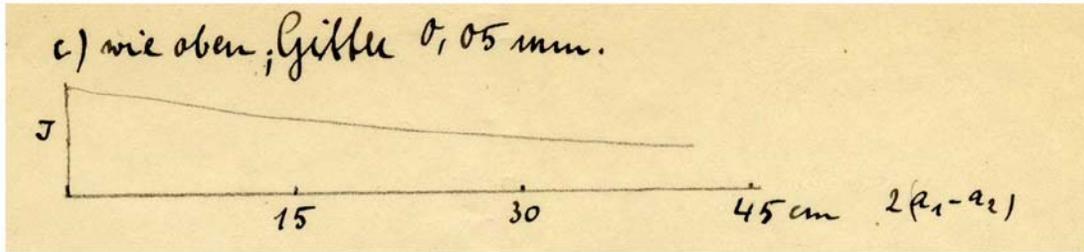

Rupp remarked that possibly the interference did not completely disappear at the minima due to diffraction at the grid. He also reported on another experiment in which both a grid (as in the Wire Grid Experiment) and a lens (as in the *Spiegeldrehversuch*) were placed in front of the interferometer—this arrangement did not have a clear outcome however.

Einstein replied to these results on May 18:

> That the experiments with the finer [i.e. the 0.05 mm] grid did not give a periodicity in the interference visibility is [...] in flagrant contradiction to the theory. The theory, after all, fully accounts for the diffraction.[60]

Indeed, for a finer grid one would expect a variability in the visibility of the interference with the minima close together. But that was not all: "The arrangement with the grid and one lens is theoretically completely obscure." Rupp had here confused the Wire Grid Experiment and *Spiegeldrehversuch*. Einstein decided that the experiments had not been conclusive at all and urged Rupp to better study his theory, because "after all, a joint publication demands a clear confrontation of theory and experiment."[61]

Rupp sent Einstein a whole new set of data just two days later. He now was sure he could "report to you results that are in agreement with the theory."[62] Rupp claimed that he had done his latest experiments independent of Einstein's last letter; after once more looking at the Einstein's theory he had decided to redo his experiments, before receiving the letter of May 18. On this occasion Rupp did find a variability in the visibility of the interference for the finer grid, apparently precisely as Einstein expected:

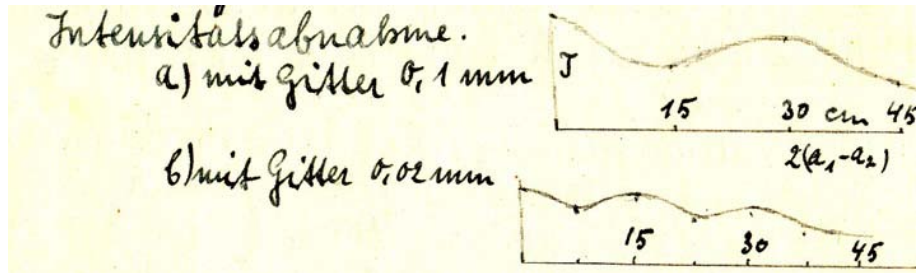

He stated that the improved results were due to placing a slit closer to the canal ray beam. According to the theory, however, this should not have any effect.[63]

Einstein responded immediately to Rupp's latest results and made clear that "the results can still not be regarded as a confirmation of the theory."[64] He had decided to check Rupp's numbers and found some alarming discrepancies:

> 1) The separation between the points of optimal interference should, for the 0.02 mm grid, be *five times* smaller than with the 0.1 mm grid. (In your experiments it is only two times smaller.)
>
> 2) The value for the separation of the optimal interference is also incorrect. With $b$ the distance between neighboring lines of the grid, $v$ the speed of the canal rays, then that separation has to be $bc/v$, i.e. in your first experiment $0.01 \times 3 \cdot 10^{10} / 1.9 \cdot 10^7 = 16\,\text{cm}$, while it is 30 cm according to your experiment. Or is in your grid this separation $2 \cdot 0.01\,\text{cm}$, as the lines and openings are 0.01 cm each? Then it is still not correct for the finer grid.[65]

Indeed Rupp had defined the values of his grids in an ambiguous fashion; should Einstein's second point hold, then all of Rupp's reported results so far would have been completely wrong—the results for the 0.02 mm grid were in any case incorrect.

Rupp responded on May 31st and set out to clear up the confusion on the values of his grids:

> The first grid contains 100 clear and dark parts in 1 cm, i.e., the distance between the bright centres is $2 \cdot 0.01\,\text{cm}$. The other grid has 200 parts/cm [Teile/cm], so the distance between two lines is $2 \cdot 0.05\,\text{cm}$ (not 0.02 cm as I inaccurately miswrote).[66]

Unfortunately, this announcement introduced its own element of ambiguity: for a grid with 100 clear and 100

dark parts in 1 cm, the distance center clear-clear would be 0.01 cm. Rupp's $2 \cdot 0.01$ cm would apply to the case in which there would be 50 clear and 50 dark parts on 1 cm. Presumably, that was then what Rupp meant and how the faulty factor of two that would affect all of his results so far was to be explained. For the second grid, Rupp made a mistake of another factor of ten and he claimed that the earlier reported value—0.02 mm, which according to Einstein had led to an incomprehensible and faulty factor of five—had just been a slip of the pen.

This was not all that Rupp had to report: he had yet another whole new set of data. He claimed that he now had used the purest canal ray light, which ought to explain yet another improvement of his results—though one wonders why he had not used this light in the first place. Nevertheless, these were his most impressive and complete results so far (see figure 9). Also, and perhaps because Einstein had now shown him exactly how to apply his relatively simple formulas (1) and (2), the values for the separation between minima and maxima of the interference visibility were in line with what one would expect on the grounds of Einstein's theory (except for yet another faulty factor of ten for the third grid). Rupp had lost quite a bit of his confidence at this point: earlier he had announced his results as "certain" (on May 14) or "in agreement with the theory" (on May 20), but now he modestly stated that it was up to Einstein to "decide how these results compare to the theory" (May 31), and the very next day he sent yet another nervous note on what the values of his grids had been, in all his prior runs.[67]

One would expect that this constant confusion would cast some doubt on the reliability of Rupp's results. Einstein wrote Rupp again on June 3rd and his reaction is both surprising and revealing:

> The experiments that you reported to me in your letter of May 31st are fully satisfying and can be considered a convincing confirmation of the theory.[68]

Einstein chose to ignore all, and settled for the last results that Rupp had sent: theory confirmed. He next hoped that Rupp would turn his attention to the *Spiegeldrehversuch*, but was already quite convinced of what the outcome was going to be:

> If the experiment with the lens also succeeds, then there is no doubt that the theory is correct; actually, that can already not be questioned.[69]

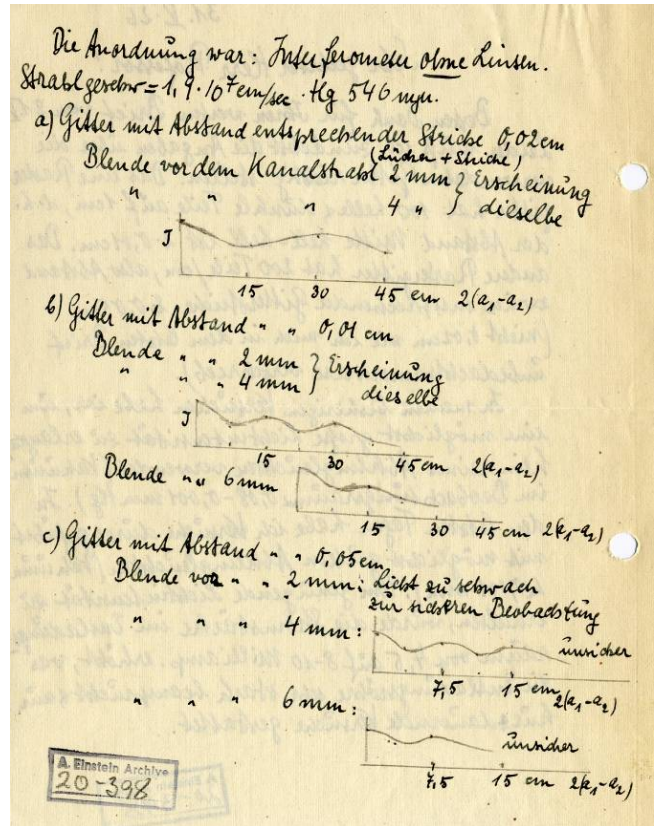

Figure 9: Second page of Emil Rupp's letter to Einstein of 31 May 1926 (EA 20 398-2).

## Rupp reports results for the Rotated Mirror Experiment

Rupp reported his results on the *Spiegeldrehversuch* on June 15; he had again used the *Hg* canal ray at 5461 Å, with speed $1.9 \cdot 10^7$ cm/s and had put a lens with a focal distance of 14 cm in front of the interferometer. The path difference in the interferometer, $d = 2(a_1 - a_2)$ with the distances defined as in figure 2, was put at 20 cm. According to Einstein's theory, maximum visibility of the interferences should then be reached with an angle between the mirrors of $\beta/2 = 4.5 \cdot 10^{-4}$ radians. Rupp had mounted the rotating mirror, mirror $S_2$ in figure 2, on a table that could be moved by turning a screw; moving the screw through 360 degrees would correspond to a rotation of the mirror of $4.3 \cdot 10^{-3}$ radians, so the screw ought to be turned through about 38 degrees to get an optimal interference pattern.

Rupp first adjusted the mirrors at optimal interference with *Hg* light from a resting source; he then turned

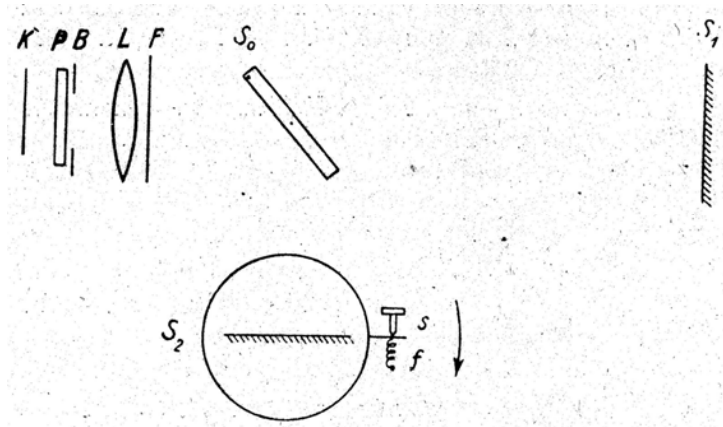

Figure 10: Rupp's diagram in Rupp 1926b (note 72) in which he drew the angle of rotation exactly opposite to what should have been observed (his mirror $S_2$ is closer to $K$, relative to $S_1$, than mirror $S_2$ in figure 8). The beam at $K$ is moving up.

on the canal ray beam:

> Interf[ences] at 0 degrees most unclear, become more clear near +15 degrees, bright and sharp between 30 and 40 degrees, then however, the clarity declines rapidly, at 45 degrees there are hardly any [fringes] visible.[70]

At first sight, Rupp seems to have clearly confirmed Einstein's prediction for the *Spiegeldrehversuch*. Furthermore, these were not just the results of only one run: Rupp ensured Einstein that "these experiments have been repeated 4-5 times during 3 nights, with the same result."[71] He had even reversed the direction of the canal ray beam, and indeed found that in that case the mirror had to be rotated in the opposite direction.

One can unequivocally see in what direction Rupp rotated his mirror when looking at the figure he presented in his publication later that year (see figure 10). Rupp made clear that in the diagram, the canal ray beam is moving upwards.[72] The mirror closest to the canal ray beam is then rotated clockwise, according to Rupp. However, that is the wrong direction of rotation: in this arrangement, the mirror should have been rotated counter-clockwise. There is an easy way to see this: with the canal ray beam moving upward, the light emitted in the top end of the canal ray beam that still reaches the interferometer is slightly red-shifted, and light emitted in the bottom end of the canal ray beam that just reaches the interferometer would be slightly blue-shifted. If this light is to interfere, that is, if the mirror rotation is to wipe out the disturbing phase differences introduced by the beam motion Doppler shifts, then the red-shifted light should travel a slightly shorter distance and the blue shifted light should travel a slightly longer distance to mirror $S_2$. With Rupp's direction of rotation however, exactly the opposite occurs. Had the rotating mirror been further away from the canal than the non-rotating mirror, as in figure 8, then this would have been the correct direction of rotation. But with the arrangement of mirrors as in Rupp's paper, his rotation went in the wrong direction.

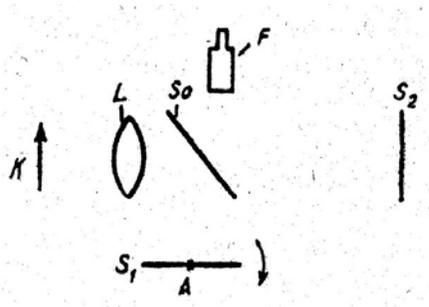

Figure 11: Figure taken from Einstein's publication in the Academy proceedings (Einstein 1926b [note 45]). The same version was contained in the manuscript version of this article that Rupp used.

So Rupp claimed to have positively carried out Einstein's *Spiegeldrehversuch*—in fact, during 3 nights, and no less than 4-5 times each night—but explicitly and repeatedly gave the incorrect direction for the rotation of the mirror. How could he have made such a mistake? It seems unavoidable to conclude that Rupp never observed the mirror rotation bring about interferences and that he just reported to have confirmed what he believed to be Einstein's correct prediction. In fact, Einstein had made the same mistake in his theoretical article, and also in its draft manuscript version that Rupp held in his hands as he was writing Einstein.[73] Certainly, Einstein's theory is without flaws, but he himself misdrew the angle of rotation in a diagram that he had included (see figure 11). Rupp repeated that mistake in his own publication, but now stated that that was what he had positively observed—repeatedly, and once more emphasized that reversing the direction of the beam also led to a reversal of the angle of rotation.[74]

In 1935, following Rupp's fall from grace and in the midst of the controversy over what elements of his work could be trusted, the Munich experimentalists Walther Gerlach and Eduard Rüchardt published a short note in the *Annalen der Physik* in which they made very clear that Rupp had confirmed Einstein's faulty diagram.[75] They even included a figure that appears to be a copy of Einstein's, but with the direction of rotation corrected (figure 12). The Munich group, in 1930, had redone Rupp's *Spiegeldrehversuch*—their work will be discussed soon—with a decidedly negative result. Gerlach and Rüchardt had published their figure partly because they had been urged to do so by the AEG's Ramsauer: the latter had believed it necessary to settle the controversy over Rupp's canal ray work once and for all.[76]

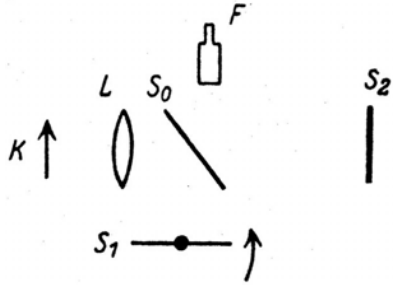

Figure 12: The corrected version as in Gerlach and Rüchardt 1935 (note 75).

## Einstein and Rupp and their joint publications

Let us return to the spring of 1926—before both Rupp's and also Einstein's second article had come out. On June 18, Atkinson's article appeared, three days after Rupp had informed Einstein of his results for the *Spiegeldrehversuch*. As discussed, this unequivocally stated that Rupp could not have observed interferences at the path differences at which he claimed to have observed them in his original *Annalenarbeit*, because of both the beam and thermal motion Doppler effect.[77]

Rupp now first told Einstein that he had, in fact, already when working on the *Annalenarbeit*, rotated a mirror "and in a completely empirical way arrived at a compensation of the Doppler effect."[78] He also made clear that he had not informed Lenard of their joint work; Einstein agreed that that was the best decision to avoid unfortunate repercussions and he further urged Rupp to send a draft of his latest work to Berlin so that it could be published in the Academy Proceedings.[79] Rupp next wrote to Einstein from his new address in Göttingen; he had sent his paper to Max Planck for publication, and added that he had:

> [...] included a paragraph on the highest attained path difference of $H_\beta$—in opposition to Atkinson in *Die Naturwissenschaften*—that shows that in my habilitation work I had unknowingly carried out the *Drehspiegelversuch* and in this way arrived at interferences at up to 15 cm path

difference.[80]

Einstein, in his reply, informed Rupp that he had asked that both of their papers would come out together, one immediately after the other. He agreed that:

> The success of your earlier experiments can certainly only be explained by an unconscious rotation of the mirror.[81]

Einstein may not yet have noticed the appearance of Atkinson's critique; of course, in the *Spiegeldrehversuch*, the Doppler effect due to the beam's motion would be corrected for by a rotation of a mirror, but this could still not cancel a Doppler effect due to the thermal motion of the atoms. Rupp had added something on this aspect in his manuscript, but Einstein, in a later letter, disagreed with what he had written:

> I have taken the liberty to strike the last sentence before the "conclusion" in your paper, as it contains an incorrect statement. A rotation of the mirror can never cancel out additional irregular, for example thermal velocities in the canal ray beam. [...] It is not at all clear how interferences with such long path differences are possible.[82]

The above strongly suggests that Einstein had by now indeed seen Atkinson's damning review—or, at least, had come to the same conclusion. Faced with even Einstein's incomprehension, Rupp must have realized that he had maneuvered himself into an untenable position and could now only agree that his earlier results were indeed quite puzzling:

> Thank you for your correction. I had been tempted to put in that sentence as I could otherwise not account for the non-disappearing of interferences of $H_\beta$ with respect to the thermal motion. [...] I am now certain that in my earlier article [i.e. the *Annalenarbeit*], the Doppler components of the canal velocity had been compensated for by an unconscious rotation of the mirror. The question is only: is it possible that the electric field or the arrangement of slits and liquid air cooling in the canal influence the thermal motion, such that sideways components are reduced? One is inclined to answer that question in the negative, whereas the result of the experiment speaks in its favor. New experiments must be done on this issue, but I wish to postpone those for the time being.[83]

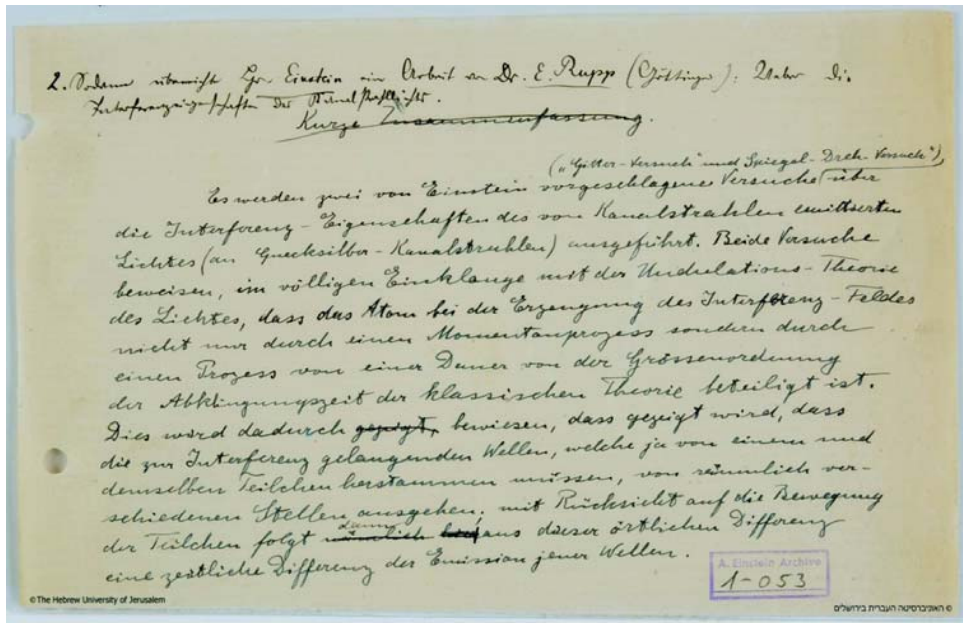

Figure 13: Abstract of Rupp's paper in Einstein's hand. The handwriting at the top of the page is by Max Planck. The image can be found at http://www.alberteinstein.info/High/01-053.jpg. See also Einstein to Emil Rupp, October 1926, EA 70 717.

Rupp stood by his observations and suggested yet other circumstances that might explain them. Did Einstein now realize that there was something rather dubious about Rupp's work? He had seen him change his data repeatedly—and each time in better accordance with his own criticism, and on one occasion in no less than two days. He had had to accept that Rupp claimed to earlier have "unknowingly" or "unconsciously" rotated a mirror, and he will likely have seen that Rupp's work was highly controversial amongst experimentalists, leading to very public criticism in *Die Naturwissenschaften*. He himself was now also convinced that, in fact, Rupp's results were incomprehensible. So, did Einstein choose to suspend the publication of Rupp's piece, so that an additional round of checks and balances could take place?

The answer is no: Rupp's paper was presented by Einstein to the Prussian Academy in a session on 21 October 1926, and it appeared in print in the Academy's proceedings in November of 1926—the articles by Einstein and Rupp came out back to back, and reprints circulated with both papers bound together, with a joint cover page that displayed both titles.[84] Einstein referred in his article to Rupp's claims[85] and he had even written the abstract of Rupp's paper (see figure 13).

Einstein's relation to experiment, for the Rupp case and in general, will be discussed in the conclusion of this article. Here, aside from that issue, let us finish by observing that Rupp's publication still stated that the 15.2 cm coherence length for the $H_\beta$ canal ray source could be understood with the mirror rotation.[86]

## Decline and Fall of Emil Rupp: The Munich Experiments

In following years, Rupp's results still received positive attention. Max von Laue, for instance, included the Einstein-Rupp experiments in a review article on the "optics of moving bodies."[87] At the same time, the number of controversies over various elements of Rupp's oeuvre grew and quite a few of his experiments were criticized. In 1928, his recent work on the polarization of canal ray light came under strong criticism by Georg Robert Döpel and Rudolf von Hirsch. In the ensuing polemic and much to the latter's surprise, Rupp actually stated in a footnote that he had superimposed and recopied photographic images to arrive at his published pictures. Of course, von Hirsch found this "inconsistent with the established principles of experimental physics."[88] In 1930, after having moved to the AEG, Rupp again produced contentious work on the Mott scattering of electrons. Allan Franklin, in his historical analysis, has argued that all of Rupp's findings on this topic had likely been forged.[89] Thirdly, Walther Gerlach and Hans Buchner took issue with Rupp's early work on the magnetic properties of phosphors. Gerlach's *Nachlass* at the Deutsches Museum in Munich contains proofs of a critical note by Gerlach and Buchner and a reply authored by Rupp. However, the journal—*Annalen der Physik*—eventually published just one short article in 1931, signed by all three authors and of a less sharp tone.[90] It may have become weary of attacks on Rupp: during 1930 Harald Straub, a doctoral student from Gerlach's group in Munich, had engaged Rupp in a polemic on the *Spiegeldrehversuch* that had consumed quite a bit of its space.

Indeed, on 4 June 1930, Straub had defended a Ph.D. dissertation at the University of Munich in which he described his unsuccessful attempt to redo the *Spiegeldrehversuch*. Wilhelm Wien, the noted authority on canal rays, had originally put him on this subject; after Wien had passed away in 1928, supervision of Straub's thesis had been taken over by Gerlach and Rüchardt[91] (the latter had already criticized Rupp's earliest canal ray work in 1926). Unlike Einstein, who had wanted a confirmation of his analysis which may have made him insufficiently critical, the Munich group was certain that Rupp's claims went against their carefully assembled knowledge of the physics of canal rays.[92] As Rupp's experiments had attracted wide attention and, as Straub pointed out, "even textbooks are referring to them,"[93] they needed to be contradicted in the strongest possible terms.

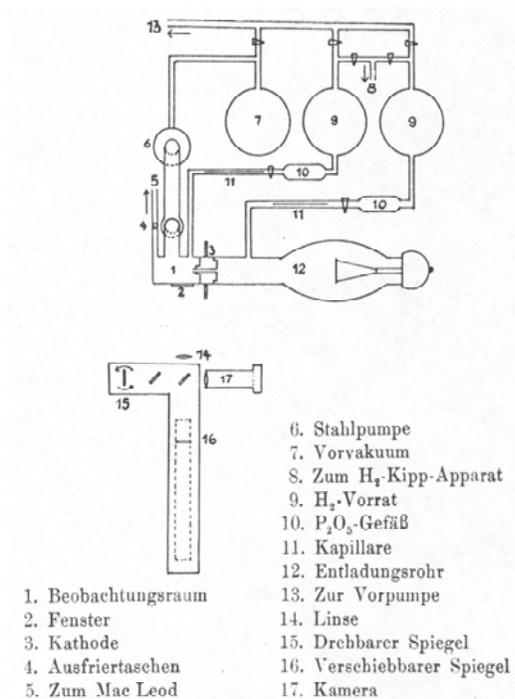

Figure 14: Straub's rotating mirror experiment (Straub 1930b [note 91]). Considerably more detail is provided for the canal ray arrangement than in Rupp's case (figure 2).

Straub's first paper in the *Annalen* left no question as to what the Munich group held of Rupp's work: it was simply impossible to observe what Rupp had claimed. Firstly, all known canal ray beams that were suited for the study of light emission were much too inhomogeneous to allow the observation of clear minima in the Wire Grid Experiment or find a sharp enough angle in the *Spiegeldrehversuch*—or observe interferences at tens of centimeters. Straub once more pointed to the line width and the associated thermal motion that should limit the coherence length, and again stated that Rupp's result of 15.2 cm for $H_\beta$ went against all reasonable expectations. He further reported the results of his own experiments (see figure 14 for his arrangement): the best he could do with a hydrogen source (actually for the $H_\alpha$-line) at rest was interferences at 4.1 cm path difference.[94] In the case of the hydrogen canal ray source, Straub could not observe any interference for whatever value of the path difference or orientation of the mirror—he did not go below 2 mm path difference.

At the same time, he measured up the velocity distribution of his canal ray source and found it so spread out that one could easily understand his negative result. With an *Hg* canal ray and without a lens or wire grid in front of the interferometer, Straub saw interferences at at most 0.7 mm path difference. He concluded that Rupp's results had now been clearly contradicted: they were simply impossible—nothing was wrong with Einstein's theory, but one needed homogeneous canal rays to be in a position to confirm it.[95]

Rupp immediately set out to respond to Straub's publication. On 12 July 1930 he sent a first draft to Einstein, to whom he also announced his intention of redoing his canal ray experiments—Straub was dismissed as a clumsy graduate student with a lousy apparatus. Einstein suggested to invite Straub once Rupp had his experiment up and running again, but cautioned him not to engage the polemic in too sharp a tone.[96] Soon Rupp—banking on the suggestive nature of visual evidence—sent out pictures of interferences to a number of people, and claimed that they were taken in the *Spiegeldrehversuch*. Hermann Mark, of the IG Farben laboratory, expressed that Rupp's pictures were "so pretty that one can hardly doubt the correctness of your claim." Rudolf Ladenburg, whom had not seen any interferences when he had visited Rupp's laboratory earlier, was now fully convinced of the veracity of Rupp's findings.[97] Max von Laue also thanked Rupp for his pictures and said that he did not find Straub's paper convincing, "as long as the exact same apparatus under the exact same conditions has not given results that deviate from yours." It is of relevance to point out that von Laue had just published a substantial paper on electron diffraction together with Rupp and thus had a continuing stake in the latter's reputation. Nevertheless, von Laue in his letter to Rupp did agree with Straub (and Atkinson) that "the thermal motion of the molecules perpendicular to the direction of the beam appears not to play a role in your case. But why not?"[98]

The reply that Rupp got from his former institute head in Göttingen will have disquieted him more: although the discussion, Robert Pohl said, was hard to follow from a distance, he indicated that the decision on a professorship that Rupp had been seeking in Danzig had been put off for half a year. Pohl had strongly recommended Rupp and now urged him to "finish off" Straub by publishing new results.[99] Straub indeed appeared to have thwarted Rupp's professorship; Franz Wolf from Danzig wrote him the following:

> Thank you very much for sending those beautiful interference pictures. I showed them to

> Professor [Eberhard] Buchwald, and he was initially very excited. But then he grew a bit suspicious, as he thought that perhaps you had somehow ended up with the wrong interferences. There are so many interferences in such a Michelson interferometer that it is easy to make a mistake. He does not believe in such long coherence lengths. Your prospects here seem to me to be not so good. People have begun to get second thoughts because of the Straub paper.[100]

For Rupp, not just the result of one experiment, nor the distinction of having successfully carried out an experiment for Einstein was at stake: his entire reputation and future career hung in the balance.

Rupp submitted his reply to Straub to the *Annalen der Physik* on August 18, 1930. However, the final article appeared in the much later issue of 10 November.[101] In the intervening months, Rupp altered his manuscript and the journal's proofs numerous times, each time after having learnt the contents of an intended retort by Straub that was shared with him prior to the publication of both papers. The Gerlach *Nachlass* in Munich contains correspondence that allows a reconstruction of some of Rupp's alterations.

After some back and forth, Gerlach sent Rupp a latest version of Straub's intended rebuttal on October 2. The Munich group had abstracted a demand from Einstein's original paper that Rupp would not have satisfied: Straub believed—incorrectly—that the distance separating the lens from the rotating mirror had to be equal to the focal length of the lens. There were in fact no grounds for such a condition in the theory, and it is not clear how or why Straub had come up with the demand; it nevertheless seemed valid to those engaged with the experiment. Straub, in his draft rebuttal, charged that Rupp had violated this 'Einstein requirement' in both the 1926 and his latest version of the *Spiegeldrehversuch*.[102]

In both Rupp's 1926 publication in the Berlin Academy Proceedings as in the reply to Straub's thesis that Rupp had just submitted to the *Annalen*, the focal length had been given as 14 cm. Rupp, behind the back of the journal's editors, managed to convince its printers to change '14' into '64' in his reply. Of course, this would undermine Straub's latest critique and the latter had to withdraw his retort. Eduard Grüneisen, editor of the *Annalen*, found out about Rupp's manoeuvre and admitted to Gerlach that he was dumbfounded by this behavior.[103] The publication of Rupp's paper was subsequently also postponed. Rupp blamed a printer's error that he had only noticed once he had seen Straub's latest version. "Of course, the Einstein requirement was always fulfilled."[104]

At Rupp's invitation,[105] Straub was to visit his laboratory on the morning of October 10 and observe him do his experiment. While in Berlin, Straub proposed a change of strategy to Gerlach: they should only react to Rupp once his paper had appeared in print. If, for instance, Rupp was now claiming that his results for hydrogen stood in "no immediate relation to the *Spiegeldrehversuch*" (whereas earlier it had been an "unconscious" mirror rotation that accounted for his huge path differences) then Straub believed it in their best interest that "he should publish that," because "then one can nail him splendidly."[106] The Munich group indeed from now on followed this tactic: Straub observed Rupp do his experiments, kept his most important objections to himself, returned to Munich and waited with writing his final critique.[107]

Gerlach had lost all patience with Rupp; he had earlier already concluded that Rupp's interference pictures were quite likely false[108] and was firmly convinced that Rupp had not complied with the "Einstein requirement" on any occasion. Rupp now emphatically argued that he had indeed fulfilled the "Einstein requirement" each and every time that he had carried out the *Spiegeldrehversuch*. In a letter to Gerlach of October 17, he included new pictures in which one clearly saw an interference pattern when the requirement was satisfied, but when $f = 40$ cm and the distance lens—rotating mirror was at 30 cm, Rupp pointed out that one did not "see any interferences, as the Einstein requirement is not fulfilled."[109]

Rupp's reply to Straub's thesis in the November issue of the *Annalen* included three pictures—two contained an interference pattern, one did not as the mirror had not been properly rotated on that occasion. He claimed to have successfully repeated the *Spiegeldrehversuch* and to have attained interferences for *Hg*-canal rays at 20 cm path difference. Rupp reproached Straub for having primarily looked at the interferences for hydrogen and not mercury—but he had repeated the hydrogen experiments too:

> [I have] attained again at about 9 cm path difference the interferences that I described earlier, and have further studied the conditions for their occurrence. It must be stressed that these experiments stand in no direct relation to the Einsteinian *Spiegeldrehversuch*. I here just state the empirical observation that it is possible to attain interferences with high path differences also with hydrogen canal rays. I add that an essential condition for this is that the compensator plate $P_2$ [*P'* in figure 2] must be at a particular angle, to be determined empirically, with the light ray.[110]

Initially Rupp had attained high path differences for hydrogen by "empirically" moving slits around, then by "empirically" rotating mirrors, but now, an "empirical" rotation of the compensator plate brought the interferences about. The focal length of the lens now came close enough to satisfying the "Einstein requirement" and Rupp emphasized that his pictures were "in agreement with the designated path difference."[111]

Straub had seen Rupp's arrangement in Berlin in the meantime, and replied in a note re-submitted in late December of 1930. Of course, he did not fail to bring to attention that in 1926 Rupp had pointed to the mirror rotation to explain the 15.2 cm path difference for $H_\beta$. Straub further had a number of serious objections regarding Rupp's set-up, but undoubtedly the most damning point of critique was that, yes, he had seen interferences with *Hg* sources, but only when sudden discharges in the tubes caused a flash—discharges during which the canal ray beam disappeared altogether. Straub was convinced he had not seen canal ray interference patterns, but interference of light from stationary sources.[112]

Things were beginning to look bad for Rupp: even Pohl felt compelled to point out to him that it was now very much "in your interest that you elucidate your observations."[113] Rupp replied once more to Straub in the *Annalen* in a note submitted on 15 January 1931: he published new pictures, and this time one could purportedly see a projection of the canal ray beam in the interference pattern, suggesting that Rupp was in fact looking at interferences of canal ray light, not just of light emitted by non-moving sources. The picture with the interference pattern was taken when the "Einstein requirement" for the distances was fulfilled, but Rupp also produced a picture for the case that this requirement was violated. In that case one did not see an interference pattern, and he argued that this established that he was looking at a moving source. Namely, should the source for the interference pattern be at rest, then the location of the lens ought not to have any effect: one should have seen interference in both pictures, regardless of whether the "Einstein requirement" had been fulfilled or not, according to Rupp. Finally, he once more emphasized that the 15.2 cm coherence length found for hydrogen had nothing to do with the *Spiegeldrehversuch*, and contended that he had held all along that this value was "theoretically incomprehensible."[114] He again suggestively pointed to a rotation of the compensator plate to account for this observation.

The Munich group had meanwhile contacted Einstein. Gerlach had met him at the 1930 Solvay conference

and in a follow-up letter of 31 October 1930, Rüchardt asked Einstein whether it was correct "that you doubt that in the "*Spiegeldrehversuch*" the condition [...] *focal length of lens = distance lens–rotating mirror* must necessarily be satisfied."[115] He pointed out that Rupp now laid great emphasis on the fulfillment of this condition; as seen, Rupp even argued that because of this condition he could show that he had a moving source. Rüchardt explained to Einstein that Straub's results were easy to understand if one only took into account that the available canal ray beams that were suited for studying light emission were much too inhomogeneous.

Einstein replied: "I believe that a precise verification of the Rupp experiments is of interest"—mildly suggesting that he too had become less convinced of the veracity of Rupp's results. "If [in Straub's] experiments only such short interference lengths can be attained, then that must be due to an inhomogeneity in the beam, as you quite rightly point out." Einstein went on to explain once more some of the theoretical considerations of the *Spiegeldrehversuch* that have been outlined here earlier, and concluded that "in the whole argument, the *position* of the lens does not matter, only its focal length."[116] So, he made unequivocally clear that there was no "Einstein requirement." This meant that Rupp should have seen interferences for any position of the lens, also if he actually did have a moving source.

The Bavarians must have been quite content with this reply. Einstein tacitly appeared to agree with them on all points and even handed them another argument to dismiss Rupp's claims with. Gerlach wrote Grüneisen a short reply to Rupp's second contra-Straub piece—in it, he expressed the hope that Einstein would soon publicly explain his position on the "Einstein requirement."[117] That hope never materialized however and the Straub-Rupp polemic ended with a single statement by Straub saying that he had nothing further to add.[118] Rupp had dug a deep enough hole for himself with his contradictory statements and spurious explanations, for all to see.

The polemic with Straub hurt Rupp's reputation badly. Furthermore, as pointed out already, it was not the only subject in which his work was heavily disputed. His correspondence from early 1932 suggests that his funding at the AEG had by that time been substantially cut and that he was looking for employment elsewhere.[119] His situation soon improved again until he had a serious fiasco at a Physical Society meeting in 1933—this time, Ramsauer terminated his appointment, but with a year's notice. In that year, Rupp produced his amazing positron production papers, upon which Ramsauer decided to re-hire him.[120] After the positron work had come out, in mid-December 1934, Lange and Brasch turned to Ramsauer and claimed that the potential differences and currents in Rupp's papers must have been lower—and that therefore Rupp's results "could not have been real."[121] As discussed, this led the AEG to conduct an internal investigation that confirmed the allegations and finally brought Rupp to admit that most of his paper had been invented by him; the house of cards finally collapsed.

The repercussions were severe. Rupp lost his position at the AEG and was urged to resign his lectureship at the Technical University in Berlin. Furthermore, he was forced to retract his work publicly and had to undergo the humiliation of Ramsauer questioning the validity of all his publications. Not surprisingly, he suffered a nervous breakdown and was admitted to a sanatorium.[122] Ramsauer soon asked Gerlach to publish a final instalment in the Straub-Rupp polemic—a request that Gerlach, together with Rüchardt, fulfilled by pointing out in the *Annalen* the faulty direction of the mirror rotation in Rupp's 1926 paper in the Berlin Academy Proceedings (see figure 12).[123]

Perhaps the most telling reaction to Rupp's downfall came from the German Physical Society (DPG). It circulated a notice to its members which, after first outlining recent developments, stated that a great number of physicists, "also abroad," had repeatedly questioned Rupp's work. Rupp had often responded by pointing to supposed misprints or by explaining "his questionable results as the consequence of previously unaccounted for effects, so that in summa the correct result would again come out." The DPG held that Rupp had consistently operated this way over a period of years and had come to the following decision:

> [T]he Business Meeting of 1935 of the German Physical Society feels the obligation to advise its members no longer to refer in articles, text books or reference books, to publications and results of publications of Herr Rupp [...]. She further demands that journals that are published in cooperation with the German Physical Society no longer accept letters or articles by Herr Rupp.[124]

All traces of Rupp were supposed to disappear from the literature and he was expelled from the professional community.

# Conclusions

In what follows some concluding observations on the exchanges between Einstein and Rupp will be offered, particularly on how the relation between theory and experiment affected their collaboration. The role of the social and political context of the Rupp case—in particular that of physics in Weimar and National Socialist Germany—is discussed first.

## Rupp's demise and the socio-political context

Rupp had been educated by an early Nazi supporter, Philipp Lenard,[125] yet derived most of his renown from his collaboration with Einstein, a prominent liberal and Jew. He had supposedly done his experiments in Lenard's laboratory in Heidelberg but his work only came to light after he had left there. This must have upset the reactionary academic circles of southern Germany—Lenard and Einstein had been open antagonists since at least as early as 1918.[126]

The familiar divide in the German physics community between Berliners and southern conservatives[127] seems to exhibit itself here too: two of Rupp's most senior collaborators, Einstein and Max von Laue, were prominent members of Berlin physics, whereas Lenard and Wilhelm Wien held important chairs in the south. Also, Einstein and von Laue were foremost theoretical physicists, whereas Lenard and Wien saw themselves as exponents of a more experimentally oriented tradition. In particular Lenard was extremely critical of the relatively new theoretical discipline and for example fully rejected relativity theory (along with its author).

One might imagine that neither Einstein nor von Laue initially took much notice or gave much weight to the controversies over Rupp's results, as they appeared to have initiated in conservative quarters—quarters that had been agitating against relativity too. However, there is no evidence on this score, and relations between von Laue and Wien appear to have been cordial. Moreover, Walther Gerlach hardly fit into this ideological divide.

On the other side, the letters that Lenard wrote to Wien, who also was editor of the *Annalen der Physik*, do reflect a sense of betrayal after the Einstein-Rupp experiments had appeared in print. Initially, Lenard was quite appreciative of Rupp's work and regretted that he was going to leave Heidelberg—he even recommended Rupp for a position in Munich.[128] But in January of 1927, Lenard truly denounced Rupp:

> You will remember that a little while ago I recommended Dr. Rupp to you, and that the latter worked

> here on interference of canal ray light. A not very satisfying publication came out in your "Annalen",[129] as Dr. Rupp (seemingly) no longer wanted to continue with the matter and as my own eyes could not see the interferences of such weak light (my eyes are not bad, but certainly not young any more). Dr Rupp has talent but he was disinclined to do thorough work [...]. My education had no more effect on him; I had wished that he go somewhere else, where he could see anew that it still has value to aim higher, which is why I recommended him to you.[130]

After having distanced himself from Rupp's *Habilitation* paper, Lenard continued with dismissing his latest work.

> Now, after Dr Rupp has been away for quite a while, a publication of his has appeared in the Berlin Acad.[131] with experiments that are supposed to have been done in the Institute here, but of which I have never heard or seen anything. Also, no one else in the Institute has seen anything, and in any case, it would again have been things that would only have been quite poorly visible. It is a continuation of the earlier interference paper, following Mr. Einstein's instructions, it is said. [...] Mr. E. may be satisfied with this work. I would not attach much value to it; then, if it would have been done properly, it could also have been demonstrated (there are more than enough younger eyes in the Institute) or at least the set-up of the apparatus could have been shown. Dr Rupp has however completely taken the latter apart—if it had ever been in a proper state at all—against my orders to leave everything standing (as I intended to still have some proper work done on canal ray interferences).[132]

The above makes clear that Lenard believed that Rupp had not in actual fact observed what he claimed to have observed. Perhaps he chose to inform Wien of this out of frustration with Rupp's new association with Einstein, but he had different intentions as well:

> I write you all this in order to ask you *to strike my name* should Rupp send his work to the *Annalen* (as he lists [my name] in the Acad. publication as that of one of his collaborators). [...] In almost every publication of Dr Rupp something turned out not to be quite correct or yet unclear and dodgy after it had appeared [...]. But he did not want any better![133]

In a next letter, Lenard called it a "horror" when "somebody does something like Dr Rupp, and 'confirms' a calculation (that is called 'theory')."[134] He emphasized that he still believed it to be of the utmost importance to do experiments on the interference properties of canal ray light—but such that "quantitatively usable results under varying experimental conditions" are attained. Furthermore, they should be done without any "preceding publicity." This last comment echoes a charge that was often levelled against Einstein by anti-relativists: both his and his theories' public and professional success was a result of a state of 'mass-hypnosis' that had been brought about by a concerted advertising campaign from the side of Einstein (and Jewish owned media).[135] In any case, Lenard stressed to Wien that "It would be such a pleasure for me if such experiments were carried out by you!"[136] Indeed, it is entirely conceivable that Wien and the Munich group started pursuing Rupp because they had been prompted to do so by Lenard.

Soon Wien would put Harald Straub to the task of redoing Rupp's experiments, which Straub then continued under Gerlach and Rüchardt. All the same, it would be a mistake to reconstruct Straub's work as exclusively a derivative of the Einstein versus Lenard history. The primary interest of the Munich group was not to dent Einstein's reputation: Straub's papers made a careful distinction between Einstein's theory and Rupp's results. The theory was taken as essentially correct. Straub emphasized that it was the fact that Rupp's results were "not in agreement with the observations of canal rays made thus far" which had prompted his research.[137] To be sure, Lenard later blamed Rupp's fraud on a taking on of "the Jew spirit, which has no respect for the truth"[138] but in the publications that came out of Munich, Einstein was spared criticism.

An even stronger indication of the a-political nature of the Munich group's interest in the Einstein-Rupp experiments came in 1938, when H. Billing published his results.[139] Billing, another junior scientist in the Munich laboratory, had also taken up Einstein's *Spiegeldrehversuch* under Gerlach and Rüchardt's supervision. He stated that with the means available in 1926 the experiment could not be carried out—but in 1938, sufficiently homogeneous and parallel canal rays had become available. Indeed, he could see interferences of the $H_\beta$-line up to some 1.5 mm path difference (that is, at two orders of magnitude less than Rupp had claimed) and found a confirmation of a central prediction of Einstein's analysis: for arrangements with two different focal distances of the lens, he reported data for the angle of rotation at varying path differences that confirmed the linear dependency in formula (7). So, the Munich group ended up confirming Einstein's theory.

Einstein and von Laue were slow in disqualifying Rupp. In 1932, well after the Rupp-Straub polemic, and following other controversies, Einstein still suggested to a collaborator, Walther Mayer, to ask Rupp to do an experiment for them.[140] Von Laue, in the same year, made efforts to ensure that Rupp would get funding through the Rockefeller foundation.[141] One could argue that as former endorsers of his work they had an interest in his reputation and were therefore less inclined to write him off. Or, their slow response could yet be taken to suggest that neither of them initially attached great value to the criticism directed at Rupp due to socio-political motivations at their end. In 1936, however, Max von Laue reported to Einstein, by then exiled in the United States, of the events at a meeting of Saxon physicists in Halle where a student of Rudolf Tomaschek—a prominent supporter of *Deutsche Physik*[142]—outlined the theory of Einstein's experiments. "This young man clearly intends to eliminate the swindle that Rupp has carried out [...] by doing his own experiments."[143] Von Laue by now actually had reservations against Einstein's analysis and brought these forward in Halle— "something extremely comical happened: Tomaschek defended you against me! That you would certainly not have expected." This episode again strongly suggests that the interest of experimentalists in Einstein's experiments, be they Nazi's or otherwise, transcended the political and social divides. By the same token, von Laue, a prominent supporter of Einstein and a critic of National-Socialist rule,[144] now just as well judged Rupp's canal ray work to be a "swindle", regardless of his own political or social allegiances. So, rather than attributing Einstein and von Laue's slow reactions to socio-political factors, there is a much more likely cause for their continued trust in Rupp's work: the theorist's prejudices when confronted with experiment.

**Einstein, Rupp and Fraud: Theory and Experiment**

Why did Einstein go along with Rupp's experimental work? Einstein was no practitioner of canal ray experiments, which obviously made judgment harder. He might also have been impressed by Rupp's affiliation with the Heidelberg laboratory, even if he did not hold Lenard's theoretical or personal qualities in high regard.[145] As French has pointed out too, Rupp usually communicated his results in a fairly convincing manner with a professional presentation that provided considerable circumstantial detail.[146] Einstein must of course have believed that Rupp was reporting his results truthfully. Yet, when looking at the Einstein-Rupp exchanges, one might still be inclined to infer that he directed Rupp to data that fitted his theory and dismissed data that did not. Other explanations seem more likely, however, as several considerations show.

Einstein obviously had an immediate interest in Rupp's results for the Wire Grid Experiment and *Spiegeldrehversuch*. Yet both Einstein's re-evaluated theory and Rupp's results seemed to counter, or at least nuance, an easy confirmation of his light quantum hypothesis—a confirmation that he initially expected and an hypothesis in which he held considerably more stock than in his analysis of the *Spiegeldrehversuch*. Secondly, some of Rupp's initial data just did not make any clear sense. They simply did not provide a good test of Einstein's theory, but rather only seemed confused, as in the conflation of the Wire Grid and Mirror Experiment. On that account it is understandable that Einstein kept suggesting to Rupp to critically re-evaluate his results. To the outsider the repeated confusion over the value of certain parameters and data may seem an indication of fraud, but to Einstein, who, as said, must have believed that Rupp was reporting actual measurements, they may rather have looked like a symptom of sloppiness and misapprehension.

On the other hand, Einstein could of course have grown suspicious about the slipshod way in which Rupp appeared to arrive at his results. Instead, Einstein was convinced, certainly by May 1926, that his theoretical analysis had to be correct, and expected Rupp only to find results that were in complete agreement with his analysis: he therefore pressed him long enough until he got the results that he expected. Once Einstein believed that Rupp had found such confirmation, he apparently felt no further need to scrutinize the latter's work—or, for that matter, to attend to the Atkinson publication. This does suggest a strong theoretical prejudice on Einstein's part.

This theoretical prejudice calls to mind the experimentalist who stops searching for systematic error in his arrangement once he gets the results that he expects on the grounds of theory or prior experimentation. To see such a theoretical prejudice play a prominent role in the Einstein-Rupp experiments is not entirely surprising, since on the only occasion that Einstein published experimental results of his own, obtained in collaboration with Wander J. de Haas in 1915, precisely such a theoretical predisposition foreclosed a search for systematic errors and made him settle for an incorrect value of the gyromagnetic ratio of the electron.[147]

An increasingly less than critical attitude with regard to experimental practice[148] may be part of Einstein's broader development during the 1910's and 1920's. In September 1925, six months before Einstein's first contacts with Rupp, Paul Ehrenfest wrote to Einstein expressing his hope that an upcoming meeting in Leiden would be the occasion for discussions between Niels Bohr and Einstein that would probe the problems of the quantum, "particularly [...] regarding experiments that you always think out on the frontier between 'waves and particles'."[149] Einstein's reply is revealing:

> I no longer think about experiments on the boundary between waves and particles; I believe that this was a vain effort. Inductive means will never get you to a sensible theory, even though I do believe truly foundational experiments, like Stern and Gerlach's and Geiger and Bothe's, can be a real help.[150]

The Einstein-Rupp experiments were of course intended to probe exactly the wave-particle frontier, and they can further be seen as prime examples of the kind of "foundational experiments" that Einstein still appreciated. The above comment, however, also suggests a certain limited regard for experimentation. Einstein seemed to imply that valuable experiments are those that yield a clear "yes" or "no" answer to a question of principle. He did not expect much from inductive searching in the realm of experience.

Much has been written on Einstein's gradual shift from an empiricist to a more rationalist position, at a distance from experiment and induction on the basis of observation. Einstein came to believe that new insight for the creative theorist was rather to be found in mathematics than in experience and pointed to his 1915 discovery of general relativity to motivate his reshaped methodology. His work gradually turned away from the kind of physics that could in principle still relate to experiment and observation, towards the mathematical abstractions of his unified field theories.[151] Klaus Hentschel has further pointed out that in the 1920's Einstein had become reluctant about getting involved in discussions about experiments as he saw himself more and more exclusively as a theorist.[152] Seen in this context it is not too surprising that he held a rather uncritical attitude with regards to Rupp's work.

As for other theorists, Max von Laue's case nicely reveals how the credit that is awarded a theory might affect the assessment of an experiment; it further suggests that the reverse may sometimes hold too. Initially, in 1928, von Laue included the Einstein-Rupp experiments in a review article on the "optics of moving bodies,"[153] and, as discussed, he chose the side of Rupp at the time of the Straub-Rupp polemic. Furthermore, he continued to support Rupp well in to 1932. But when by 1936 he had become convinced that Rupp had carried out a "swindle", he not only believed that the experiments had been fabricated, he also began to think that their theory was incorrect, and started urging Einstein to publish a withdrawal of his analysis.[154] To be sure, the objections he reported to Einstein were serious enough for a renewed discussion between them[155]—yet, one cannot help but notice an awkward reversal. Von Laue's 1936 response, despite that his new theoretical objections were perhaps not entirely unfounded, does suggest that the belief in the quality of an experiment can on occasion strongly influence the assessment of a theory as well.

By 1936 Einstein had reached the point that he defended his arguments—against von Laue's new objections—without even once mentioning the name of Emil Rupp. Einstein wrote in reply to von Laue:

> I do not consider my considerations of those days to be superfluous or false. I even believe that they still are fairly interesting. Because in my opinion, we today still lack a theory that can be taken serious. Pardon my putting it in such detail. But I see that you have not appreciated the point that makes my considerations of those days meaningful. Of course, also back then they did not require any confirmation by experiment.[156]

To Einstein in 1936 the theory was obviously correct, so no experiment had ever been necessary: where earlier his theoretical convictions had prevented him from scrutinizing Rupp, their self-evidence now allowed him—perhaps partly in embarrassment—to repress the experiments altogether. The canal ray experiments had been formulated as cases in which "our [theoretical] knowledge would make a decision possible, even without carrying out an experiment."[157] Indeed Emil Rupp's name and the Einstein-Rupp experiments, with the help of the physics community's discomfiture and the DPG's circular, gradually disappeared from view.

* J.vanDongen@phys.uu.nl. I wish to thank Diana and Jed Buchwald, Cathryn Carson, Dennis Dieks, Wilhelm Füßl, Dieter Hoffmann, Michel Janssen, David Kaiser, A.J. Kox, Christoph Lehner, Jürgen Renn, Tilman Sauer, Urs Schoepflin, Suman Seth, Jos Uffink, and Barbara Wolff for help and insights that contributed greatly to this article. I am most grateful to the Max Planck Institute for the History of Science in Berlin for support and for making available to me its Rupp correspondence; I thank the Einstein Archive at Hebrew University and Princeton University Press for permission to use the Einstein correspondence in this article, and the Deutsches Museum in Munich and its staff for help and permissions to use its material. The author gratefully

acknowledges support by a Veni-grant of the Netherlands Organization for Scientific Research (NWO).

The following abbreviations are used: EA, Albert Einstein Archive, Hebrew University, Jerusalem; MPIWG, Rupp correspondence folder, Library of the Max Planck Institute for the History of Science, Berlin; GN, Walther Gerlach Papers (NL 080/124); WN, Wilhelm Wien Papers (NL 056/vorl. Nr. 008), Archive of the Deutsches Museum, Munich; *AdP*, *Annalen der Physik*; *NW*, *Die Naturwissenschaften*; *PZ*, *Physikalische Zeitschrift*; *SBPAW*, Preussische Akademie der Wissenschaften. Physikalisch-mathematische Klasse, *Sitzungsberichte*; *ZfP*, *Zeitschrift für Physik*.

[1] Walther Gerlach, Archive for History of Quantum Physics, interview with T.S. Kuhn on 18 Feb 1963, 30-31; microfilm viewed at the Deutsches Museum, Munich. The author is grateful to Suman Seth for drawing his attention to this passage.

[2] The retracted publications were: Emil Rupp, "Polarisation der Elektronen an freien Atomen," *ZfP*, *88* (1934), 242-246; "Polarisation der Elektronen in magnetischen Feldern," *ZfP*, *90* (1934), 166-176; "Versuche mit künstlich erzeugten Positronen," *ZfP*, *92* (1934), 485-512; "Versuche mit künstlich erzeugten Positronen," *Zeitschrift für technische Physik*, *15* (1934), 575-579; "Die Messung hoher Spannungen mittels Elektronenbeugung," *AdP*, *20* (1934), 594-600.

[3] Emil Rupp, "Mitteilung," *ZfP*, *95* (1935), 851. A slightly expanded version can be found in: Dr. Emil Freiherr von Gebsattel. "Abschrift. Ärztliches Gutachten." 20 Mar 1935, EA 20 430.

[4] "Bericht über die Vorgänge bei der Aufklärung der Arbeitsweise von E. Rupp im Hinblick auf seine Publikationen „Versuche mit künstlich erzeugten Positronen." ZS. f. Phys. 92 S. 485; vgl. auch ZS. f. techn. Phys. 15 S. 575." EA 20 425; the authors of this document were presumably Arno Brasch and Fritz Lange. See also Ramsauer's circular to: Johannes Stark, Max von Laue, Gustav Hertz, Richard Becker, Philipp Lenard, Max Wien, Arnold Sommerfeld, Jonathan Zenneck, Friedrich Krüger, Robert Pohl, Peter Debye, Walther Kossel, Karl Scheel, Karl Mey, Hans Rukop, Richard Fellinger, E. Hochheim, Ernst Rutherford, Irving Langmuir, 14 Jan 1935, EA 20 425.

[5] Carl Ramsauer, "Mitteilung," *ZfP*, *96* (1935), 278; Rupp (ref. 3).

[6] Werner Heisenberg, *Die Physikalische Prinzipien der Quantentheorie* (Leipzig, 1930), 59-60.

[7] A.P. French, "The strange case of Emil Rupp," *Physics in perspective*, *1* (1999), 3-21.

[8] One cannot easily underestimate its effect: see e.g. L. Bohnenkamp to Rupp: "Thank you for sending the reprints, that I study carefully. Hopefully you will not immediately become *Geheimrätlich* because of the new symbiosis with the *persona gratissima* Einstein." 16 Dec 1926, MPIWG.

[9] On the role of replication in science, see Allan Franklin, *The neglect of experiment* (Cambridge, 1986), 242-243.

[10] Emil Rupp, "Lebenslauf." EA 20 428; Emil Rupp, *Über erregende Absorption und Tilgung der Phosphore*, Ph.D. Dissertation, University of Heidelberg, 1922.

[11] Emil Rupp, "Interferenzuntersuchungen an Kanalstrahlen," *AdP*, *79* (1926), 1-34.

[12] For more on the early history of canal rays, see Henk de Regt, "Eugen Goldstein, Wilhelm Wien en de kanaalstralen, 1886-1912," *Gewina*, *12* (1989), 110-129.

[13] For an overview of Wien's work and the physics of canal rays, see Wilhelm Wien, "Kanalstrahlen," Wilhelm Wien and Friedrich Harms, eds., *Handbuch der Experimentalphysik*, *14* (Leipzig, 1927), 435-788; see also e.g. Eduard Rüchardt, "Kanalstrahlen," Rudolf Dittler, Georg Joos and Eugen Korschelt, eds., *Handwörterbuch der Naturwissenschaften*, 2nd. edn. (Jena, 1934), 754-767.

[14] Wilhelm Wien, "Über das Leuchten der Kanalstrahlen," *AdP*, *70* (1923), 1-31, and Chapter 10, "Allgemeine Theorie des Leuchtens der Kanalstrahlen," Wien (ref. 13), 732-746.

[15] Wilhelm Wien, "Über das Leuchten der Kanalstrahlen bei hohen Drucken und die Frage der Verweilzeit," *AdP*, *76* (1925), 109-123; Rupp (ref. 11), 3, 21-22.

[16] On the Michelson interferometer, see e.g. Grant E. Fowles, *Introduction to modern optics* (New York, 1989), 63-66; on its historical uses, see Michel Janssen and John Stachel, *The optics and electrodynamics of moving bodies* (Berlin, 2004: Max Planck Institute for the History of Science, Preprint 265), 15-21.

[17] Rupp (ref. 11), 6.

[18] Rupp (ref. 11), 9-11.

[19] Fowles (ref. 16), 68-74.

[20] Rupp wrote for the line width $b \sim \lambda\sqrt{\theta/m}$ with $\theta$ the temperature and $m$ for the atomic mass; Rupp (ref. 11), 9.

[21] Rupp (ref. 11), 10; when defining $\varphi$ Rupp only speaks of a lens "*L*".

[22] According to Rupp 65 cm was the largest value for the maximum coherence length that was certain (see Rupp [ref. 11], 1); he indeed referred to Lummer and Gehrcke (Otto Lummer and Ernst Gehrcke, "Ueber die Interferenz des Lichtes bei mehr als zwei Millionen Wellenlängen Gangunterschied," *Verhandlungen der Deutschen Physikalischen Gesellschaft*, *4* [1902], 337-346), who in fact however claimed even longer path differences. These claims where later withdrawn, see: Henri Buisson and Charles Fabry, "La largeur des raies spectrales et la théorie cinétique des gaz," *Journal de physique*, *2* (1912), 442-464, on 448.

[23] For this theoretical value, see Buisson and Fabry (ref. 22) and Rowan d'E. Atkinson, "Über Interferenz von Kanalstrahlenlicht," *NW*, *14* (1926), 599-600. Buisson and Fabry had measured a value similar to this for the maximum coherence length for the $H_\alpha$ Balmer line when using a Geissler tube as source.

[24] Rupp (ref. 11), 12-20.

[25] Rupp (ref. 11), 22, 33.

[26] See Robert Pohl to Emil Rupp, 11 Apr 1926, MPIWG.

[27] Eduard Rüchardt, "E. Rupp. Interferenzuntersuchungen an Kanalstrahlen," *Physikalische Berichte*, 7 (1926), 1523-1524. Upon seeing Rüchardt's note, Rupp wrote him a letter claiming that, despite what the figure suggested, the pump was actually not immediately located at the canal (Rupp's letter is mentioned in another review by Rüchardt ["E. Rupp. Zur Lichtdauer der Atome. Abklingung bei den Alkalien und im Magnetfeld," *Physikalische Berichte*, 7 (1926), 2052]). Rüchardt, although willing to correct his review, was not quite convinced. In a reply to Rupp he stated that he could not imagine how the latter had managed to maintain coherence in the hydrogen beam long enough to establish a coherence length of 15.2 cm; he argued that Martin C. Johnson of the University of Birmingham had found a much broader spectral line for $H_\beta$ than Rupp's coherence length allowed for (E. Rüchardt to E. Rupp, 7 Oct 1926, MPIWG; Rüchardt referred in this letter to Atkinson [ref. 23] and Martin C. Johnson, "The distribution of intensity in a positive ray spectral line," *Proceedings of the Physical Society of London*, 38 [1925], 324-334; see also "Part II." Ibid., 39 [1926], 26-38.)

[28] Walter Grotrian, "Interferenzuntersuchungen an Kanalstrahlen," *NW*, 14 (1926), 399-400.

[29] Albert Einstein, "Vorschlag zu einem die Natur des elementaren Strahlungs-Emissionsprozesses betreffenden Experiment," *NW*, 14 (1926), 300-301. Recent literature on Einstein's light quantum includes: David C. Cassidy, "Einstein and the quantum hypothesis," *AdP* (Leipzig), 14 Supplement (2005), 15-22; Robert Rynasiewicz and Jürgen Renn, "The turning point for Einstein's *annus mirabilis*," *Studies in history and philosophy of modern physics*, 37 (2006), 5-35.

[30] Albert Einstein, "Über ein den Elementarprozess der Lichtemission betreffendes Experiment," *SBPAW*, (1921), 882-883; reprinted in Michel Janssen, Robert Schulmann, József Illy, Christoph Lehner, Diana Kormos Buchwald, Daniel Kennefick, A.J. Kox and David Rowe, eds., *The collected papers of Albert Einstein. Volume 7, the Berlin years: Writings, 1918-1921* (Princeton, 2002), 483-487.

[31] Albert Einstein, "Theorie der Lichtfortpflanzung in dispergierenden Medien." *SBPAW*, (1922), 18-22. For a historical discussion, see Martin J. Klein, "The first phase of the Bohr-Einstein dialogue," *Historical studies in the physical sciences*, 2 (1970), 1-39, and the annotation of Document 68 in Janssen et al. (ref. 30).

[32] For Einstein's response to the Compton experiment, see Albert Einstein, "Das Komptonsche Experiment," *Berliner Tageblatt*, 20 Apr 1924, Supplement I.

[33] Walther Bothe and Hans Geiger, "Über das Wesen des Comptoneffektes; Ein experimenteller Beitrag zur Theorie der Strahlung," *ZfP*, 32 (1925), 639-663; Albert Einstein, "Observações sobre a situação atual da teoria da luz." *Revista da academia Brasileira de sciencias*, 1 (1925), 1-3. See also: Anthony Duncan and Michel Janssen, "On the verge of Umdeutung in Minnesota: Van Vleck and the correspondence principle (part one)", preprint, 2006: http://uk.arxiv.org/abs/physics/0610192; Abraham Pais, "Interlude: the BKS proposal," *'Subtle is the Lord...' The science and the life of Albert Einstein* (Oxford, 1982), 416-422.

[34] Walther Bothe, "Über die Kopplung zwischen elementaren Strahlungsvorgängen," *ZfP*, *37* (1926), 547-567. For Einstein's enthusiasm for Bothe's results, see Einstein to Hendrik Antoon Lorentz, 13 Mar 1926, EA 16 595.

[35] Our account here differs from that in Don Howard, "'Nicht sein kann was nicht sein darf,' or the prehistory of EPR, 1909-1935: Einstein's early worries about the quantum mechanics of composite systems," Arthur I. Miller, ed., *Sixty-two years of uncertainty: historical, philosophical and physical inquiries into the foundations of quantum mechanics* (New York, 1990), 61-109, on 85; see however also Alexander Bach, "Eine Fehlinterpretation mit Folgen: Albert Einstein und der Welle-Teilchen Dualismus," *Archive for history of exact sciences*, *40* (1989), 173-206, on 184-185.

[36] Note that Einstein's *b* is obviously different from Rupp's use of that symbol in figure 2.

[37] As in Einstein (ref. 29), 301.

[38] Einstein to Erwin Schrödinger, 26 Apr 1926, EA 22 018.

[39] Paul Ehrenfest to Einstein, 7 Apr 1926, EA 10 134.

[40] Georg Joos to Einstein, 25 Apr 1926, EA 13 164, EA 13 165. Georg Joos, "Modulation und Fourieranalyse im sichtbaren Spektralbereich," *PZ*, *27* (1926), 401-402; see also the letters that Joos wrote to Einstein, 28 Apr 1926, EA 13 466; 9 May 1926, EA 13 467.

[41] Atkinson (ref. 23); he based his expectation for the coherence length on the theoretical analysis of Buisson and Fabry (ref. 22).

[42] Einstein to Emil Rupp, 20 Mar 1926, EA 70 701.

[43] Emil Rupp to Einstein, 23 Mar 1926, EA 20 386.

[44] See Einstein to Emil Rupp, 31 Mar 1926, EA 70 702. For more on Lenard, see e.g. Charlotte Schönbeck, "Albert Einstein und Philipp Lenard. Antipoden in Physik und Zeitgeschichte," *Schriften der mathematisch–naturwissenschaftliche Klasse der Heidelberger Akademie der Wissenschaften*, *8* (2000), 1-42; Alan D. Beyerchen, *Scientists under Hitler. Politics and the physics community in the Third Reich* (New Haven and London, 1977); Klaus Hentschel, ed., *Physics and National Socialism. An anthology of primary sources* (Basel, Boston, Berlin, 1996).

[45] Albert Einstein, "Über die Interferenzeigenschaften des durch Kanalstrahlen emittierten Lichtes." *SBPAW*, (1926), 334-340. The article was presented in the session of 8 Jul 1926 (see Einstein's abstract in the proceedings of that session [*SBPAW*, (1926), 274]) and published in the proceedings of the session of 21 Oct 1926.

[46] See Einstein to Emil Rupp, 9 May 1926, EA 70 706; Einstein (ref. 45).

[47] For a derivation, see Miles V. Klein and Thomas E. Furtak, *Optics*, 2nd. ed. (New York, 1986), on 284-307.

[48] Einstein to E. Rupp, 31 Mar 1926, EA 70 702. Emphasis in original.

[49] In fact, this idea was already outlined by Einstein as early as 1922: see "Nederlandsche Natuurkundige Vereeniging. Vergadering op zaterdag 29 april 1922 in het natuurkundig laboratorium der Universiteit van Amsterdam." *Physica. Nederlandsch*

*tijdschrift voor natuurkunde*, *2* (1922), 158-159. In that seminar, Einstein briefly suggested rotating the mirror of an interferometer in front of a canal ray source. The question this was to address was indeed the duration of light emission. However, Einstein did not initially recognize his 1922 idea in Rupp's experiment, as exhibited by his initial incomprehension regarding how Rupp could have cancelled the relevant Doppler effect. By 1926 he may well have forgotten his earlier suggestion; a strong indication for this is that Einstein explained the idea of the mirror rotation again to Ehrenfest in 1926 without a reference to the seminar in Amsterdam (see his letter of 12 Apr 1926, EA 10 136), even though Ehrenfest had participated in the discussion there. It further seems that there is no mention of the 1922 seminar in any of Einstein's 1926 correspondence on the canal ray experiments— particularly not in his letters to Ehrenfest or Rupp. In 1922 the idea appears to only have come up on this occasion, suggesting that at that time it was of minor concern to Einstein.

[50] Einstein to Rupp (ref. 48), Einstein (ref. 45); on its earlier occurrence in 1922, see note 49. In his analysis Einstein once more first put the source at infinity, again allowed because of the symmetry in the arrangement and the particular kind of interferometer used.

[51] Einstein to Paul Ehrenfest, 12 Apr 1926, EA 10 135.

[52] The letter to Ehrenfest of 12 April (ref. 51) suggests that Einstein was then already expecting a classical result for the canal ray experiments. In a later letter to Erwin Schrödinger (26 Apr 1926, EA 22 018), Einstein still expressed that he believed that the particulate picture was going to be confirmed. On 5 May he however clearly spoke out again in favour of the classical result (Einstein to Emil Rupp, 5 May 1926, EA 70 705).

[53] Einstein to Emil Rupp, 18 Apr 1926, EA 70 703; Emil Rupp to Einstein, 11 Apr 1926, EA 20 389.

[54] A limited number of the arguments they exchanged are presented here, but these give a fairly complete and representative account of the dynamic.

[55] Emil Rupp to Einstein, 29 Apr 1926, EA 20 392.

[56] Emil Rupp to Einstein, 1 May 1926, EA 20 393.

[57] Einstein to Emil Rupp, 5 May 1926, EA 70 705; For Rupp emphasizing the role of his slits, see Emil Rupp to Einstein, 29 Apr 1926, EA 20 392.

[58] Einstein to Emil Rupp, 9 May 1926, EA 70 706. The manuscript of Einstein's publication (ref. 45) has survived and can be viewed online; see http://www.alberteinstein.info, Document 1-54.00.

[59] Emil Rupp to Einstein, 14 May 1926, EA 20 395.

[60] Einstein to Emil Rupp, 18 May 1926, EA 70 707.

[61] Ibid.

[62] Emil Rupp to Einstein, 20 May 1926, EA 20 397.

[63] In Einstein's theory, the beam's velocity distribution was taken to be completely homogeneous; the relocation of the slit (*B* in figure 2) should therefore not have any effect. Rupp hinted that as the canal ray atoms move further away from the entry canal, their velocity distribution would however be less homogeneous because of interactions. The slit in front of the canal ray beam would limit the light coming from the endpoint of the beam (see Emil Rupp to Einstein, 20 May 1926, EA 20 397). Einstein, in his reply, however reminded Rupp that in his original paper (ref. 11), the latter had claimed that interactions with stationary atoms would not be an issue (Einstein to Emil Rupp, 21 May 1926, EA 70 708).

[64] Einstein to Emil Rupp, 21 May 1926, EA 70 708.

[65] Ibid.

[66] Emil Rupp to Einstein, 31 May 1926, EA 20 398.

[67] Emil Rupp to Einstein, 1 June 1926, EA 20 399.

[68] Einstein to Emil Rupp, 3 June 1926, EA 70 709.

[69] Ibid.

[70] See Emil Rupp to Einstein, 15 June 1926, EA 20 400.

[71] Ibid.

[72] Emil Rupp, "Über die Interferenzeigenschaften des Kanalstrahllichtes," *SBPAW*, (1926), 341-351, on 348-349. The article was presented by Einstein and appeared in the proceedings of the session of 21 Oct 1926.

[73] For that manuscript, see ref. 58; its page 4 contains Einstein's diagram.

[74] Rupp (ref. 72), 348-349.

[75] Walther Gerlach and Eduard Rüchardt, "Über die Kohärenzlänge des von Kanalstrahlen emittierten Lichtes," *AdP*, *24* (1935), 124-126.

[76] Carl Ramsauer to Walther Gerlach, 19 Aug 1935, GN 124-01; an inventory of the Gerlach Papers is available: Wilhelm Füßl, ed., "Der wissenschaftliche Nachlaß von Walther Gerlach (1889-1979)," *Veröffentlichungen aus dem Archiv des Deutschen Museums*, *3* (1998), 2 vols.

[77] Atkinson (ref. 23).

[78] Emil Rupp to Einstein, 29 June 1926, EA 20 401.

[79] Einstein to Emil Rupp, 2 Jul 1926, EA 70 711.

[80] Emil Rupp to Einstein, 21 Aug 1926, EA 20 402.

[81] Einstein to Emil Rupp, 27 Aug 1926, EA 70 712.

[82] Einstein to Emil Rupp, 5 Nov 1926, EA 70 715.

[83] Emil Rupp to Einstein, 8 Nov 1926, EA 20 409.

[84] A copy of such a reprint can be found in the Gerlach *Nachlass*, GN 124-03.

[85] See Einstein (ref. 45), 340.

[86] Rupp (ref. 72), 349-350.

[87] Max von Laue, "Die Optik der bewegten Körper," Wilhelm Wien and Friedrich Harms, eds., *Handbuch der Experimentalphysik*, *18* (Leipzig, 1928), 38-104, on 47-48, 58-60.

[88] Emil Rupp, "Zur Polarisation des Kanalstrahllichtes in schwachen elektrischen Feldern. I. Das Abklingleuchten von $H_\beta$ in einem elektrischen Querfeld," *AdP*, *85* (1928), 515-528; Rudolf von Hirsch and Robert Döpel, "Die „Axialität" der Lichtemission und verwandte Fragen," *PZ*, *29* (1928), 394-398; Emil Rupp, "Bemerkungen zu R. v. Hirsch u. R. Döpel: „Die Axialität der Lichtemission und verwandte Fragen,"" *PZ*, *29* (1928), 625-626; Rudolph von Hirsch, "Erwiderung," *PZ*, *29* (1928), 730-731, on 731. An overview of criticism of Rupp's work on canal rays during the years 1926-1928 can be found in the Gerlach *Nachlass*: "Bemerkungen zu Ruppschen Kanalstrahlarbeiten," GN 124-01; a more elaborate discussion of this episode is contained in French (ref. 7), 11-12.

[89] Franklin (ref. 9), 227-229. See also Olivier Darrigol, "A history of the question: Can free electrons be polarized?" *Historical studies in the physical sciences*, *15* (1984), 39-79, on 67-74, and French (ref. 7), 12-13.

[90] The paper under attack was: Emil Rupp, "Über das magnetische Verhalten der Phosphore," *AdP*, *78* (1925), 505-557. The proofs are entitled: "Kritische Bemerkungen zur Abhandlung von Hrn. E. Rupp: „Über das magnetische Verhalten der Phosphore." Von Hans Buchner und Walther Gerlach."—"Erwiderung zu H. Buchner und W. Gerlach. Von E. Rupp." The typesetting is that of the *Annalen der Physik*; see GN 124-03. The final publication is: Hans Buchner, Walther Gerlach and Emil Rupp, "Bemerkungen zu der Abhandlung von E. Rupp „Über das magnetische Verhalten der Phosphore,"" *AdP*, *8* (1931), 874-876.

[91] According to Walther Gerlach, "Erinnerungen an Albert Einstein 1908-1930," Peter Aichelburg and Roman Sexl, eds., *Albert Einstein: sein Einfluss auf Physik, Philosophie und Politik* (Braunschweig, 1979), 199-210; Straub's thesis is: Harald Straub, *Über die Kohärenzlange des von Kanalstrahlen emittierten Leuchtens*, Ph.D. Dissertation, University of Munich, 1930. His results were also published in the *Annalen der Physik*: Harald Straub, "Über die Kohärenzlange des von Kanalstrahlen emittierten Leuchtens," *AdP*, *5* (1930), 644-656.

[92] Wien's 1927 *Handbuch* review of the subject (ref. 13) ominously did not include references to Rupp's work.

[93] As in Straub, *Annalen* (ref. 91), 647; examples would be the texts of Heisenberg (ref. 6) and von Laue (ref. 87).

[94] This was an improvement of over 10% on the accepted value in the literature, indicating that he had a good optical arrangement.

[95] See Straub, *Annalen* (ref. 91).

[96] Emil Rupp to Einstein, 12 Jul 1930, EA 20 416; Einstein to Emil Rupp, 17 Jul 1930, EA 20 418.

[97] Hermann Mark to Emil Rupp, 23 Aug 1930, MPIWG; Rudolf Ladenburg to Emil Rupp, 2 Sep 1930, MPIWG.

[98] Max von Laue to Emil Rupp, 28 Aug 1930, MPIWG; Max von Laue and Emil Rupp, "Über Elektronenbeugung an nichtmetallischen Einkristallen," *AdP*, *4* (1930), 1097-1120. This paper actually also contained an observation that von Laue soon found puzzling ("sehr sonderbar") and in need of further clarification; see his attempt at the latter: Max von Laue, "Notiz zur Dynamik der Randschichten eines Kristalls vom NaCl-Typus. Zusatz zu der vorhergehenden Arbeit von M. v. Laue und E. Rupp," *AdP*, *4* (1930), 1121-1132.

[99] Robert Pohl to Emil Rupp, 19 Jul 1930, MPIWG.

[100] Franz Wolf to Emil Rupp, 21 Aug 1930, MPIWG.

[101] Emil Rupp, "Erwiderung zu der Dissertation von H. Straub „Über die Kohärenzlänge des von Kanalstrahlen emittierten Lichtes,"" *AdP*, *7* (1930), 381-384.

[102] Straub's second draft is entitled "Über die Kohärenzlänge des von Kanalstrahlen emittierten Leuchtens. Bemerkung zu der Erwiderung von E. Rupp." It is contained in GN 124-01; Walther Gerlach to Emil Rupp, 2 Oct 1930, GN 124-01.

[103] Eduard Grüneisen to Walther Gerlach, 6 Oct 1930, GN 124-01.

[104] Emil Rupp to Walther Gerlach, 5 Oct 1930; Emil Rupp to Walter Gerlach, 8 Oct 1930, GN 124-02.

[105] Ibid.

[106] "Das soll er mal ruhig veröffentlichen, da kann man ihn ja herrlich festnageln!" Harald Straub to Walther Gerlach, 9 Oct 1930, GN 124-02.

[107] Harald Straub to Walther Gerlach, 10 Oct 1930, GN 124-02. Much to Gerlach's chagrin, Rupp had managed to write a skewed review of Straub's original paper for the *Physikalische Berichte*, the German Physical Society's abstract publication (Emil Rupp, "Harald Straub. Über die Kohärenzlänge des von Kanalstrahlen emittierten Leuchtens," *Physikalische Berichte*, *11* [1930], 2154). Gerlach corrected the situation by having his own review published, and made sure that the index only referred to his note (Eduard Rüchardt to Karl Scheel, 18 Oct 1930; Karl Scheel to Walther Gerlach, 25 Oct 1930; Walther Gerlach to Karl Scheel, 31 Oct 1930, GN 124-02; Walther Gerlach, "Harald Straub. Über die Kohärenzlänge des von Kanalstrahlen emittierten Leuchtens," *Physikalische Berichte*, *11* [1930], 2783).

[108] Earlier Rupp had also sent Gerlach pictures of interference patters; the Munich group measured these up and concluded that Rupp could not have observed interferences at 20 cm path difference, but at 8 to 10 mm at best. Rupp next claimed that he had sent magnifications and that his originals did comply with the larger path difference (Rupp to Gerlach, 16 Sep 1930; Gerlach to Rupp, 23 Sep 1930; Rupp to Gerlach, 24 Sep 1930; Gerlach to Eduard Grüneisen, 27 Sep 1930, GN 124-01).

[109] Emil Rupp to Walther Gerlach, 17 Oct 1930; Gerlach to Rupp, 14 Oct 1930, GN 124-02.

[110] As in Rupp (ref. 101), 384.

[111] Ibid.

[112] Harald Straub, "Über die Kohärenzlänge des von Kanalstrahlen emittierten Leuchtens. (Bemerkung zu der Erwiderung von E. Rupp)," *AdP*, *8* (1931), 286-292; the Gerlach Nachlass contains a letter from Straub to Gerlach (10 Oct 1930, GN 124-02) in which Straub states that he has drafted a protocol of observations done in Rupp's lab. The protocol would state that interferences were only visible during discharges, and would have been signed by Straub, the neutral observer Marianus Czerny, and Emil Rupp.

[113] Robert Pohl to Emil Rupp, 13 Jan 1931, MPIWG.

[114] Emil Rupp, "Antwort auf die vorstehende Bemerkung von H. Straub," *AdP*, *8* (1931), 293-296, on 295.

[115] Eduard Rüchardt to Einstein, 31 Oct 1930, EA 20 422.

[116] Einstein to Eduard Rüchardt, 12 Nov 1930, GN 124-02.

[117] Walther Gerlach to Eduard Grüneisen, 30 Jan 1931, GN 124-02.

[118] See Straub, *Annalen* (ref. 112): Straub's statement of 2 Feb 1931 is found on its final page 292. He did publish another paper showing that the homogeneity of the *Hg* canal ray beam did not improve for lower pressures: Harald Straub, "Die Breite des Dopplerstreifens im inhomogenen Kanalstrahl," *AdP*, *402* (1931), 670-671.

[119] See letters by Arthur H. Compton to Emil Rupp, 6 Feb 1932, MPIWG; James Franck to Emil Rupp, 25 Jan 1932, MPIWG. However, in 1932 the AEG laboratory faced major budget cuts due to the economic depression; its staffing dropped from 75 to 44 that year. See Detlef Lorenz, *Das AEG-Forschungsinstitut in Berlin-Reinickendorf. Daten, Fakten, Namen zu seiner Geschichte 1928-1989* (Berlin, 2004), on 12.

[120] Robert Pohl to Emil Rupp, 11 Feb 1932, MPIWG; Ramsauer to Stark et al. (ref. 4).

[121] Lange and Brasch, according to Ramsauer in his letter to Stark et al. (ref. 4); for references to Rupp's positron papers, see note 2.

[122] Rupp (ref. 3), Ramsauer (ref. 5); the last according to Ramsauer, note 4.

[123] For Ramsauer to Gerlach, see note 76; Gerlach and Rüchardt 1935 (ref. 75).

[124] Undated typed document in the Gerlach Nachlass, GN 124-01. On the document is written in Gerlach's hand: "Mitt. der Geschäftsversammlung der DPG 1935 an ihre Mitgl."

[125] Lenard actually attended a rally for Adolf Hitler on 15 May 1926, right in the middle of the Einstein-Rupp collaboration (see Beyerchen [ref. 44], 97).

[126] On the antagonism between Einstein and Lenard, see e.g. Schönbeck (ref. 44) and the editorial note "Einstein's encounters with German anti-Relativists," in Janssen et al. (ref. 30), 101-113.

[127] For instance in 1920, during discussions on restructuring of the DPG, see Paul Forman, "Il Naturforscherversammlung a Nauheim del settembre 1920: una introduzione alla vita scientifica nella repubblica di Weimar," Giovanni Battimelli, Michelangelo de Maria and Arcangelo Rossi, eds., *La ristrutturazione delle scienze tra le due guerre mondiali* (Rome, 1984), 59-78. Another contentious issue between Berliners and southern physicists was the creation of the journal *Zeitschrift für Physik*; see e.g. Einstein to Arnold Sommerfeld, 18 Dec 1919, Doc. 219 in Diana Kormos Buchwald, Robert Schulmann, József Illy, Daniel Kennefick, Tilman Sauer, Virginia Iris Holmes, A.J. Kox and Ze'ev Rosenkranz, eds., *The collected papers of Albert Einstein. Volume 9, the Berlin years: Correspondence January 1919-April 1920* (Princeton, 2004), 309-310; for an overview, see Paul Forman, "The financial support and political alignment of physicists in Weimar Germany," *Minerva*, *12* (1974), 39-66.

[128] He however also said that Rupp would still benefit from "supervision." Philipp Lenard to Wilhelm Wien, 14 June 1926, WN. See also Lenard to Wien, 19 Apr 1926, WN.

[129] I.e. Rupp (ref. 11).

[130] Philipp Lenard to Wilhelm Wien, 9 Jan 1927, WN.

[131] I.e. Rupp (ref. 72).

[132] See note 130.

[133] Ibid.

[134] Philipp Lenard to Wilhelm Wien, 23 Feb 1927, WN.

[135] Jeroen van Dongen, "Reactionaries and Einstein's fame: "German scientists for the preservation of pure science," relativity and the Bad Nauheim conference," *Physics in perspective*, *9* (2007), 212-230; Hubert Goenner, "The reaction to relativity theory I: The anti-Einstein campaign in Germany in 1920," *Science in context*, *6* (1993), 107-133.

[136] See note 134.

[137] See Straub, *Annalen* (ref. 91), 646.

[138] Philipp Lenard, "Erinnerungen eines Naturwissenschaftlers, der Kaiserreich, Judenherrschaft und Hitler erlebt hat." Typescript, Heidelberg University, on p. 78. Quoted excerpt dated at 1930-1931, but according to its author, later additions had been made.

[139] H. Billing, "Ein Interferenzversuch mit dem Lichte eines Kanalstrahles," *AdP*, *32* (1938), 577-592.

[140] Jeroen van Dongen, "Einstein's methodology, semivectors and the unification of electrons and protons," *Archive for history of exact sciences*, *58* (2004), 219-254, discusses Einstein's collaboration with Mayer. For the reference to Rupp, see Einstein to Walther Mayer, 10 or 11 Aug 1932, EA 18 130.

[141] See Max von Laue to Emil Rupp, 24 Feb 1932, MPIWG.

[142] For more on *Deutsche Physik*, or "Aryan physics," and its opposition to Einstein's theories, see Beyerchen (ref. 44).

[143] Max von Laue to Einstein, 8 June 1936, EA 16 110.

[144] See Beyerchen (ref. 44), 64-66.

[145] On Lenard, see Einstein to Max Born, 26 Oct 1920, Doc. 182 in Diana Kormos Buchwald, Tilman Sauer, Ze'ev Rosenkranz, József Illy, Virginia Iris Holmes, Jeroen van Dongen, Daniel Kennefick and A.J. Kox, eds., *The collected papers of Albert Einstein. Volume 10, the Berlin years: Correspondence May-December 1920 and supplementary correspondence 1909-1920* (Princeton, 2006), 468-469; Albert Einstein, "Meine Antwort. Ueber die anti-relativitätstheoretische G.m.b.H," *Berliner Tageblatt*, 27 Aug 1920 (Morning Edition), 1-2; Doc. 45 in Janssen et al. (ref. 30), 345-347.

[146] French (ref. 7), 13.

[147] Albert Einstein and Wander J. de Haas, "Experimenteller Nachweis der Ampèreschen Molekularströme." *Verhandlungen der Deutsche Physikalische Gesellschaft*, *17* (1915), 152-170; "Experimental proof of the existence of Ampère's molecular currents," *Proceedings of the Koninklijke Akademie van Wetenschappen te Amsterdam. Section of the sciences*, *18* (1915-16), 696-711 (Docs. 13 and 14 in A.J. Kox, Martin J. Klein and Robert Schulmann, *The collected papers of Albert Einstein. Volume 6, the Berlin years: Writings 1914-1917* [Princeton, 1996], 150-189; see also its editorial note "Einstein on Ampère's molecular currents," pp. 145-149). For the historical discussion of these experiments, see Peter Galison, *How experiments end* (Chicago, 1987), 21-74.

[148] An interesting parallel may be drawn between the case of Rupp and Einstein's lenient attitude regarding the observational work on relativity of the astronomer Erwin Freundlich, who was also charged with serious scientific misconduct: see Klaus Hentschel, *The Einstein tower. An intertexture of dynamic construction, relativity theory, and astronomy* (Stanford, 1997).

[149] Paul Ehrenfest to Einstein, 16 Sep 1925, EA 10 110.

[150] Einstein to Paul Ehrenfest, 18 Sep 1925, EA 10 111.

[151] On this shift, see Gerald Holton, "Mach, Einstein and the search for reality," *Thematic origins of scientific thought: Kepler to Einstein* (Cambridge MA, 1973), 219-259; John Norton, "'Nature is the realization of the mathematically simplest ideas': Einstein and the canon of mathematical simplicity," *Studies in history and philosophy of modern physics*, *31* (2000), 135-170; Jeroen van Dongen, *Einstein's unification: General relativity and the quest for mathematical naturalness*, Ph.D. Dissertation, University of Amsterdam, 2002. On the discovery of general relativity, see: Jürgen Renn, ed., *The genesis of general relativity* (4 vols., Dordrecht, 2007); on its relation to Einstein's unified field theory program, see also Michel Janssen and Jürgen Renn, "Untying the knot: How Einstein found his way back to field equations discarded in the Zurich notebook," Michel Janssen, John Norton, Jürgen Renn, Tilman Sauer and John Stachel, *The Zurich notebook and the genesis of general relativity* (Dordrecht, 2007), 839-925.

[152] Klaus Hentschel, "Einstein's attitude towards experiments: Testing relativity theory, 1907-1927," *Studies in history and philosophy of modern physics*, *23* (1992), 593-624, on 612-613. See however on his involvement with experimental physics in Leiden: Tilman Sauer, "Einstein and the early theory of superconductivity, 1917-1922," *Archive for history of exact sciences*, *61* (2007), 159-211.

[153] Von Laue (ref. 87).

[154] See note 143.

[155] Von Laue first held that nothing new could be learnt from the Einstein-Rupp experiments as the results of the experiments could be completely predicted on the basis of Maxwell theory and the special theory of relativity (ref. 143). In a second letter, he questioned however whether it was justified to presuppose an equivalence between an arrangement with the canal ray at infinity and an arrangement with the canal ray at a finite distance from the interferometer; see Max von Laue to Einstein, 29 Jul 1936, EA 16 111.

[156] Einstein to Max von Laue, 29 Aug 1936, EA 16 113.

[157] Ibid.